\documentclass[conference]{IEEEtran}
\IEEEoverridecommandlockouts

\usepackage{graphicx} 
\usepackage{bm}
\usepackage{mathrsfs}
\usepackage{amsthm}
\usepackage{amsmath}
\usepackage{amsfonts}
\usepackage[ruled,vlined]{algorithm2e}
\usepackage{algorithmic}
\usepackage{bm}
\usepackage{subfigure}
\usepackage{multirow}
\usepackage{url}

\usepackage{array}

\newcommand{\ie}{{\sl i.e.}}
\newcommand{\eg}{{\sl e.g.}}

\newcommand{\wrt}{{\sl w.r.t. }}

\newcommand{\tabincell}[2]{\begin{tabular}{@{}#1@{}}#2\end{tabular}}

\AtBeginDocument{%
  \providecommand\BibTeX{{%
    \normalfont B\kern-0.5em{\scshape i\kern-0.25em b}\kern-0.8em\TeX}}}

\begin{document}

\title{\textsc{Perfect}: A Hyperbolic Embedding for Joint User and Community Alignment}



\author{
\IEEEauthorblockN{Li Sun$^{1}$, Zhongbao Zhang$^{1\dagger}$\thanks{$^\dagger$ Corresponding Author}, Jiawei Zhang$^{2}$, Feiyang Wang$^{1}$, Yang Du$^{1}$, Sen Su$^{1}$, and Philip S. Yu$^{3}$}
\IEEEauthorblockA{
\textit{$^{1}$School of Computer Science, Beijing University of Posts and Telecommunications, Beijing, China} \\
\textit{$^{2}$IFM Lab, Department of Computer Science, Florida State University, FL, USA} \\
\textit{$^{3}$Department of Computer Science, University of Illinois at Chicago, IL, USA} \\
\{l.sun, zhongbaozb, fywang, duyang, susen\}@bupt.edu.cn, jiawei@ifmlab.org, psyu@uic.edu}
}



\maketitle




\begin{abstract}
Social network alignment shows fundamental importance in a wide spectrum of applications.
To the best of our knowledge, existing studies mainly focus on network alignment at the individual user level,
requiring abundant common information between shared individual users. 
For the networks that cannot meet such requirements, social community structures actually provide complementary and critical information at a slightly coarse-grained level,
alignment of which will provide additional information for user alignment. 
In turn, user alignment also reveals more clues for community alignment.
Hence, 
in this paper, we introduce the problem of \emph{joint social network alignment}, which aims to align users and communities across social networks simultaneously.
Key challenges lie in that (1) how to learn the representations of both users and communities, and (2) how to make user alignment and community  alignment benefit from each other.
To address these challenges, we first elaborate on the characteristics of real-world networks with the notion of \emph{$\delta-$hyperbolicity},
and show the superiority of hyperbolic space for representing social networks.
Then, we present a novel hyperbolic embedding approach for the joint social network alignment, referred to as \textsc{Perfect}, in a unified optimization. 
Extensive experiments on real-world datasets show the superiority of \textsc{Perfect} in both user alignment and community alignment.
\end{abstract}







\section{Introduction}\label{sec:intro}

Nowadays, people join in multiple social networks to enjoy more diverse services. 
The alignment across these social networks benefits a wide range of applications, such as link prediction and information diffusion \cite{zhang2016partial}, 
and thus receives an increasing attention \cite{zhang2015cosnet,mu2016user,kong2013inferring,zafarani2013connecting}. 
To the best of our knowledge, existing studies mainly focus on the alignment at the individual user level,
requiring abundant common information between shared individual users.
However, for the networks that cannot meet such requirements, 
community structure plays an important role in understanding users' social patterns.
Community alignment enriches the information across networks especially when individual users don't have enough common information for alignment.
As illustrated in Fig. 1, compared to (a), we collect more aligned users in (c) with the additional knowledge of community alignment in (b).
Furthermore, user alignment naturally reveals more clues for inferring community alignment, as shown in Fig. 1 (a) and (b).
That is, user alignment and community alignment are strongly correlated.
Hence, we rethink that: \emph{can we jointly align users and communities across different social networks?} 

\begin{figure}
\centering
     \vspace{-0.053in}
    \includegraphics[width=1.03\linewidth]{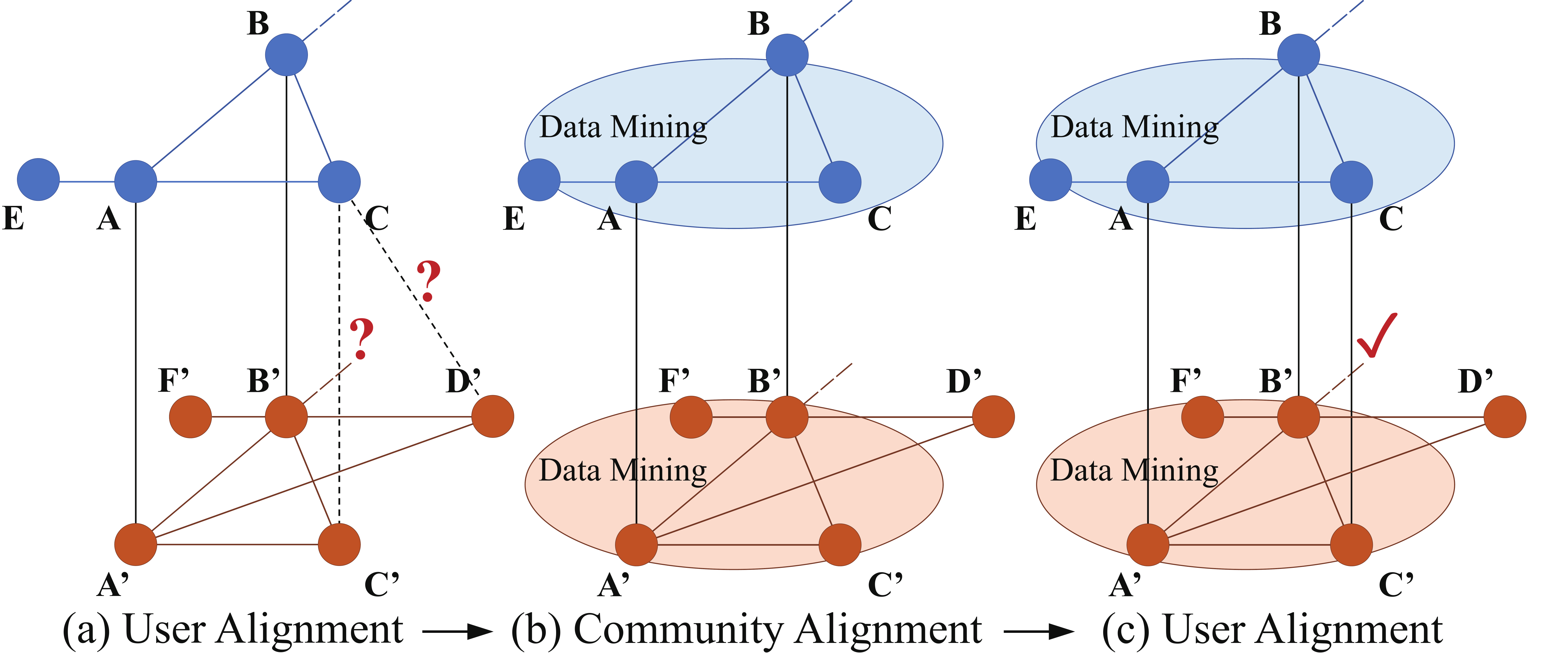}
        \vspace{-0.275in}
        \caption{Joint social network alignment: 
        Different networks are shown in different colors. 
        Black lines link aligned users. 
        Community members are grouped in the disk.
        Community alignment helps user alignment.
        It is impossible to distinguish the counterpart of C between C' and D' with network structure in (a). 
        Further knowing both C between C' are users of data mining community in (b), C is more likely to be aligned with C' rather than D' in (c). 
        In turn, aligned users across communities facilitate community alignment.
        }
           \vspace{-0.195in}
    \label{illu}
\end{figure}

To this end, we introduce the problem of \emph{joint social network alignment} in this paper. 
It is facing following challenges:
\begin{itemize}
\item \emph{How to learn the representations of both users and communities in an appropriate embedding space?} 
Accurate alignment is possible only if embeddings can capture faithful information. 
Existing methods for network alignment explicitly or implicitly work with Euclidean space \cite{man2016predict,mu2016user,zhou2018deeplink}.
However, Euclidean space tends to render reconstruction error when embedding real-world social networks \cite{chami2019hyperbolic}.
Hence, it calls for a promising embedding space for both users and communities.
\item \emph{How to make user alignment and community alignment benefit from each other?} 
To our knowledge, user alignment is widely studied while community alignment has rarely been touched before.
Though user alignment and community alignment are strongly correlated as shown in Fig. 1, 
it still remains open to make user alignment and community alignment benefit from each other.
\end{itemize}


To address these challenges, in this paper, 
we propose a novel unified hy\underline{PER}bolic embedding approach \underline{F}or the joint us\underline{E}r and \underline{C}ommunity alignmen\underline{T}, referred to as \textsc{Perfect}. 
Its essential novelty lies in that we for the first time close the loop of  user alignment and community alignment so that they benefit from each other in a unified optimization.

\begin{figure} 
\centering 
\vspace{-0.12in}
\subfigure[Zachary’s karate club network]{
\includegraphics[width=0.77\linewidth]{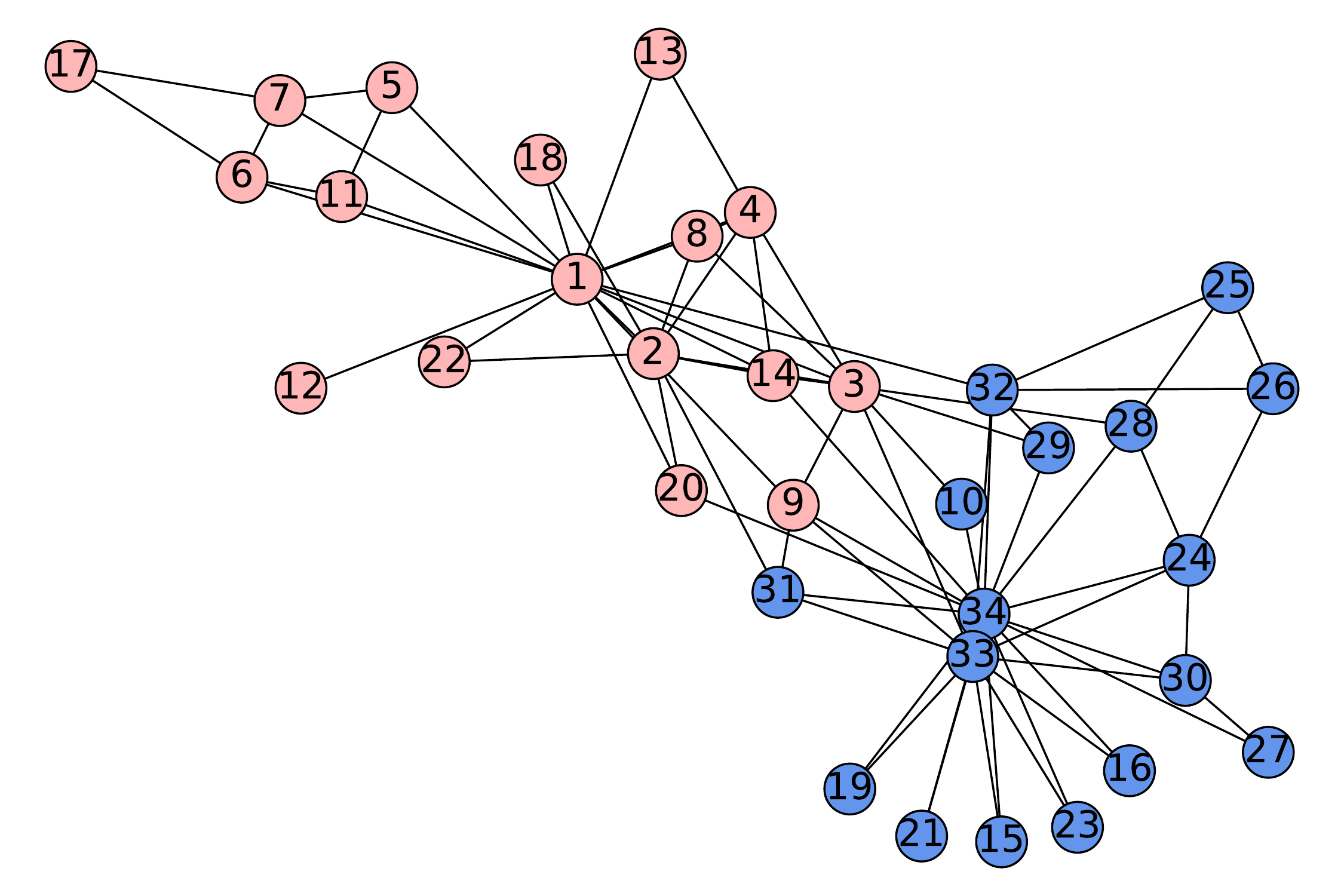}}
\vfill
\vspace{-0.12in}
\subfigure[Hyperbolic Embedding]{
\includegraphics[width=0.44\linewidth]{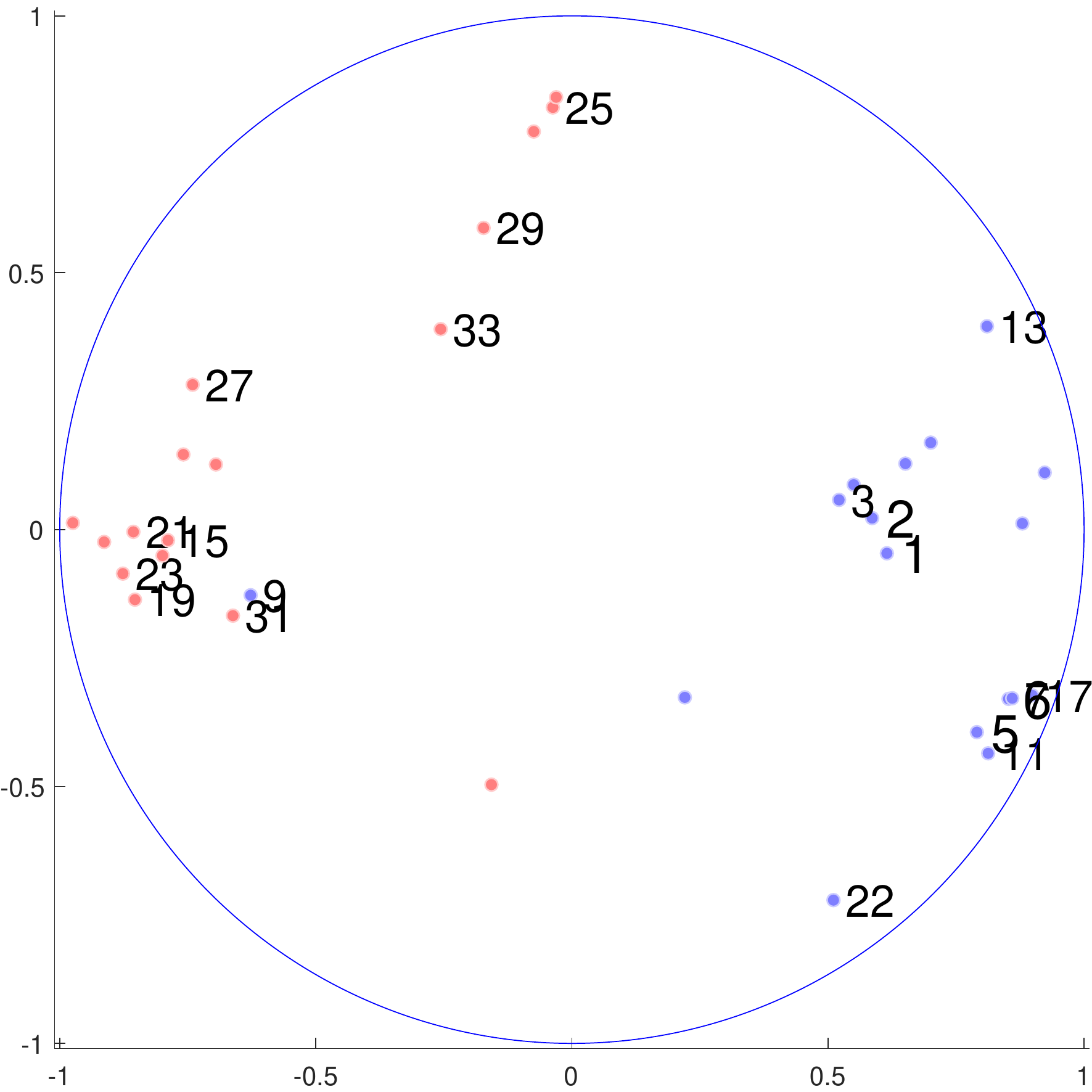}}
\hspace{0.05\linewidth}
\subfigure[Euclidean Embedding]{
\includegraphics[width=0.44\linewidth]{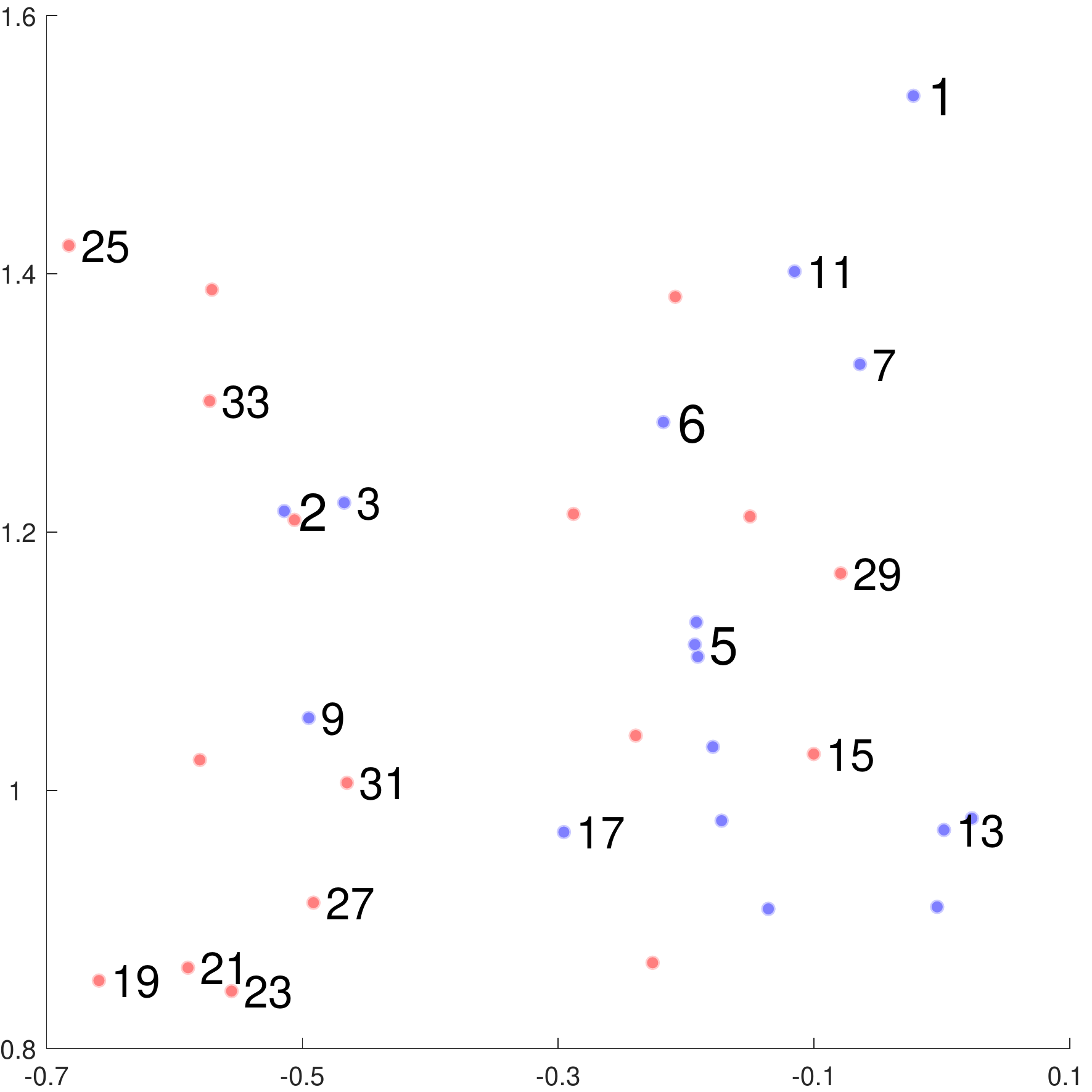}}
\vspace{-0.1in}
\caption{We embed the Zachary karate club network in (a) via the proposed hyperbolic embedding approach and the corresponding Euclidean one in $2D$ space with the same experimental settings, shown in (b) and (c), respectively. 
In (b), nodes of high degree, e.g., nodes $1$ and $2$, reside close to the origin and nodes of low degree, e.g., nodes $5$ and $6$, are positioned close to the boundary of the disk, revealing the latent hierarchy, while this does not hold in the Euclidean space in (c).
Moreover, communities are separable in (b).
}
\vspace{-0.2in}
\label{example}
\end{figure}

To address the first challenge mentioned above, we work with hyperbolic space.
To elaborate on the choice of representation space, we first give a toy example in Fig. \ref{example}:
we embed the example network in (a) with both a hyperbolic embedding approach and the corresponding Euclidean one in $2D$ space, whose results are shown in (b) and (c), respectively. 
Compared against Euclidean space, hyperbolic space tends to present the latent hierarchies among nodes in the input network.
We observe that \emph{the hierarchical characteristic} is common for social networks \cite{ravasz2003hierarchical} and, more importantly, has shown to be crucial for user alignment \cite{zhang2015cosnet} and benefits community discovery \cite{abbe2017community}. 
Furthermore, we demonstrate the hierarchical characteristic of several real-world networks with the metric of Gromovs \emph{$\delta$-hyperbolicity} \cite{chami2019hyperbolic,chen2013hyperbolicity}.
Fortunately, the hyperbolic space is well-suited to embed the latent hierarchical structures \cite{nickel2017poincare,wang2019hyperbolic,ganea2018hyperbolic}.
Thus, in \textsc{Perfect}, we embed both users and communities of each network in hyperbolic space.
We then construct a common hyperbolic subspace, 
and finally formulate the unified optimization to jointly align users and communities across social networks.

Meanwhile, to address the second challenge,  we propose an alternating Riemannian optimization algorithm so that user embeddings and community embeddings are mutually refined in the common hyperbolic subspace for the joint alignment.
Specifically, we update community embeddings and user embeddings in an alternating approach.
When updating community embeddings, 
we incorporate the knowledge of user embeddings via expectation-maximization in Riemannian manifold for community alignment.
On the other hand, when updating user embeddings, 
we incorporate the knowledge of community embeddings by conducting exponential map with Riemannian gradient for user alignment.
Furthermore, we give solid theoretical analyses on the proposed algorithm. 




Finally, we summarize the key contributions as follows:
\begin{itemize}
\item To  our knowledge, this is the first attempt to jointly align users and communities across social networks.
\item To this end, we work with hyperbolic space and propose a novel hyperbolic embedding approach with a unified optimization, \textsc{Perfect}, closing the loop of community alignment and user alignment. 
\item To address this optimization, in \textsc{Perfect}, we design a novel Riemannian alternating optimization algorithm with solid theoretical analyses.
\item We empirically evaluate the hyperbolicity of several real-world social networks and show the superiority of \textsc{Perfect}. Our code is available at \url{https://github.com/NetAligner/perfect}.
\end{itemize}


\section{Problem Statement}\label{sec:prob}

\noindent \textbf{Notations}: We use lowercase $x$, bold lowercase  $\bm x$ and bold uppercase $\bm X$ to denote scalar, vector and matrix, respectively. $\bm x^T$ denotes the transpose of $\bm x$. $|| \cdot||$ and $\langle \cdot, \cdot \rangle$ denote the usual Euclidean norm and inner product throughout this paper.

A social network is described as $\mathcal G= (\mathcal V, \mathcal E)$, where $\mathcal V=\{(v_i)\}$ is the user set of size $N=|\mathcal V|$ and 
$\mathcal E=\{(v_i,v_j)\}$ is the edge set.
A community is a subset of users $\mathcal C_p \subset \mathcal V$ with the same community label,
where $\mathcal C_p  \bigcap \mathcal C_q=\O$ for
 any $\mathcal C_p$ and $\mathcal C_q$, and $\bigcup\nolimits_{p}\mathcal C_p =\mathcal V$.
We consider a pair of social networks: $\mathcal G^s$ is the source network and $\mathcal G^t$ is the target network.
We use superscript $x$ to indicate variables associated with $\mathcal G^x$, $x \in \{s, t\}$. 
We use subscripts $i$, $j$, $k$ and $n$ to denote indexes of the users, and subscripts $p$ and $q$ to denote indexes of the communities.
The source and target networks are linked by anchor users, whose definition is given as follows:
\vspace{-0.03in}
\newtheorem*{AU}{Definition 1 (Anchor User)} 
\begin{AU}
The user who has accounts $v^s_i$  in the source network  $\mathcal G^s$  and $v^t_k$ in the target network $\mathcal G^t$  is termed as anchor user, and $(v^s_i, v^t_k)$ is called an anchor link.
\end{AU}
The set of anchor users known in advance between $\mathcal{G}^t$ and $\mathcal{G}^s$ is referred to as $\mathcal A$, which can be collected from user profiles or third-party platforms.
\vspace{-0.05in}
\newtheorem*{AC}{Definition 2 (Anchor Community)} 
\begin{AC}
Community $\mathcal C^s_p$ discovered from $\mathcal G^s$ and community $\mathcal C^t_q$ discovered from $\mathcal G^t$ are said to be anchor community iff at least $\tau$ proportion of the users in them are anchor users connecting $\mathcal C^s_p$ and $\mathcal C^t_q$.
\end{AC}
For instance, between $\mathcal C_p^s$ and $\mathcal C_q^t$, the anchor links existing among the users in them can be denoted as $\mathcal{A}_{p,q} \subset \mathcal{A}$. Then, we have $\tau = $$2|\mathcal{A}_{p,q}| /(|\mathcal C_p^s| + |\mathcal C_q^t| )$. 
Now, we formally define the problem of joint social network alignment as follows:
\newtheorem*{Prob}{The Problem of Joint Social Network Alignment} 
\begin{Prob}
Given a pair of social networks $\mathcal{G}^s$ and $\mathcal{G}^t$ with the anchor user set $\mathcal A$, 
the aim of joint social network alignment is to identify:

(1) all anchor users $\{(v^s_i, v^t_k)\}$ (\ie, user alignment)  and 

(2) all anchor communities $\{(\mathcal C^s_p, \mathcal C^t_q)\}$ (\ie, community alignment) simultaneously between the pair of social networks. 
\end{Prob}



\section{Perfect: Model}\label{sec:model}
To address this problem, we propose a novel hyperbolic embedding approach, \textsc{Perfect}, with a unified optimization.
We first elaborate on why we work with hyperbolic space, 
and then embed both users and communities in this embedding space.
Finally, we introduce the formulation of the unified optimization so that 
community embeddings and user embeddings are mutually refined for the joint social network alignment. 

\subsection{Why hyperbolic embedding?} \label{sec:why}

Here, we explain why hyperbolic space is a promising embedding space.
Recall the example in Fig. \ref{example}. 
Compared against the Euclidean space, embeddings in hyperbolic space encode the latent hierarchy among the users, \ie, users of higher centrality tend to reside closer to the origin. Such \emph{hierarchical characteristic} is of significance.

For real-world graphs, the study \cite{ravasz2003hierarchical} elaborates on the formation of their hierarchical characteristic in general.
More importantly, the hierarchical characteristic has shown to be crucial for user alignment \cite{zhang2015cosnet} and benefits community discovery \cite{abbe2017community}. 
Additionally, we demonstrate the latent hierarchy on several real-world graphs.
In order to measure the hierarchical characteristic quantitatively, we introduce the Gromov \emph{$\delta-$hyperbolicity} \cite{chen2013hyperbolicity,chami2019hyperbolic}, a metric from geometric group theory.
Note that, a lower value of $\delta$ indicates a better hierarchical structure.
Specifically, we use the following datasets:
\begin{itemize}
\item \emph{Zachary karate dataset}: The example network in Fig. \ref{example}.
\item \emph{Twitter-Quora dataset}: We collect two friendship networks, Twitter and Quora, linked by common users.
We use users' registered affiliations as community labels. 
\item \emph{DBLP-AMiner dataset}: DBLP and AMiner \cite{zhang2015cosnet} are two coauthor networks linked by common authors. 
Authors' research areas denote their community labels. 
\end{itemize}
The statistics are given in Table 1 and the corresponding \emph{$\delta-$hyperbolicity} shows their latent hierarchy.

Can we incorporate such latent hierarchy? Fortunately, we find that \emph{hyperbolic space} is well suited to embed graphs with latent hierarchical structure \cite{nickel2017poincare,wang2019hyperbolic,ganea2018hyperbolic}.
Let's take an extreme example, the tree.
We give the fact that the  $\delta-$hyperbolicity of a tree is $0$. 
The  $\delta-$hyperbolicity of the hyperbolic space (Poincar\'{e} ball model in Section \ref{sec:ball}) is $\log(1+\sqrt2)$, 
while $\delta=\infty$ for Euclidean spaces \cite{chen2013hyperbolicity}.
It is obvious that hyperbolic space better matches the $\delta$ of a tree than the Euclid.
Indeed, any tree can be embedded in a $2$-dimensional Poincar\'{e} ball with arbitrary low reconstruction error \cite{tifrea2018poincar}, while this is not true for Euclidean spaces even when an unbounded dimension is allowed. 
All these facts motivate us to leverage hyperbolic space as the representation space.
Furthermore, we will examine the effects of $\delta-$hyperbolicity with experimental results in Section \ref{sec:exp-com} and \ref{sec:exp-user}.






\subsection{The Poincar\'{e} ball Model of Hyperbolic Space}\label{sec:ball}
Now, we introduce the preliminaries of hyperbolic space for our work. 
The hyperbolic space is a kind of isotropic space with constant negative curvature,
and there are several models proposed for reasoning in hyperbolic space \cite{krioukov2010hyperbolic}.
We prefer to work with the widely used Poincar\'{e} ball model  owing to its conformality (angle-preserving \wrt Euclidean space) and convenient parameterization \cite{nickel2017poincare,wang2019hyperbolic,ganea2018hyperbolic,lxlHyperRecomm}.

The Poincar\'{e} ball model of dimension $d$ is formally defined as a smooth manifold $\mathcal B^d=\{\bm x \in \mathbb R^d| \ ||\bm x||<1\}$
endowed with a Riemannian metric:
\begin{equation}
\mathfrak{g}^{\mathcal B}(\bm{x})=\left(\frac{2}{1-\|\bm{x}\|^{2}}\right)^{2} \mathfrak{g}^{E}=\left(\lambda_{\bm{x}}\right)^{2} \mathfrak{g}^{E},
\end{equation}
which is a collection of inner product in the tangent space $T_x\mathcal B^d$ of  $\bm x \in \mathcal B^d$.
$\lambda_{\boldsymbol{x}}=\frac{2}{1-\|\boldsymbol{x}\|^{2}}$ is the conformal factor, 
and $\mathfrak{g}^{E}$ is the Euclidean metric tensor under usual Cartesian coordinates of $\mathbb R^d$. 
The Poincar\'{e} ball is conformal but wrapped (distance curving) \wrt the Euclidean space.
The distance between two points \cite{ganea2018hyperbolic}, \eg, $\bm x, \bm y \in \mathcal B^d$, is given as follows:
\begin{equation}
d(\bm x,\bm y)=\cosh^{-1}\left(1+\frac{2||\bm x - \bm y||^2}{(1-||\bm x||^2)(1-||\bm y||^2)}\right).
\label{hdist}
\end{equation}
We derive its partial derivatives \wrt $\bm y$ as follows:
\begin{equation}
\frac{\partial d(\boldsymbol{x}, \boldsymbol{y})}{\partial \boldsymbol{y}}=\frac{4 }{\beta \sqrt{\gamma^{2}-1}}\left(\frac{\|\boldsymbol{y}\|^{2}-2\langle\boldsymbol{y}, \boldsymbol{x}\rangle+ 1}{\alpha^{2}} \boldsymbol{y}-\frac{\boldsymbol{x}}{\alpha}\right),
\label{pdofd}
\end{equation}
where $\alpha=1-\|\boldsymbol{y}\|^{2}$, $\beta=1-\|\boldsymbol{x}\|^{2}$ and 
$\gamma=1+\frac{2}{\alpha \beta}\|\boldsymbol{y}-\boldsymbol{x}\|^{2}$.

\begin{center}
  \begin{table}
    \centering
      \caption{Statistics and $\delta-$hyperbolicity of the datasets}
       \vspace{-0.07in}
         \small
    \begin{tabular}{c | c c c c | c}
     \hline
  \textbf{Network }& \textbf{\#(Node)} &   \textbf{\#(Comm.)} &   \textbf{\#(Link)} &  \textbf{\#(Anchor)} & \bm{$\delta$} \\
    \hline
    Zachary & $34$                  &  $2$            &$78 $                            & - & $1$\\
      \hline
      Twitter  &        $19,438$                 &    $60$      &     $201,063$              & \multirow{2}{*}{$10,232$}& $3.5$ \\
      Quora  &           $10,638$                 &    $60$     &        $46,969$                       &                                     &  $4$\\
       \hline
      DBLP          &    $13,211$   &   $12$        &       $46,278$         &   \multirow{2}{*}{$12,213$}&  $2.5$\\
      AMiner        &      $13,213$ &      $12$      &       $46,189$             &                              & $3$\\
      \hline
    \end{tabular} 
    \vspace{-0.12in}
     \label{stat}
  \end{table}
\end{center}

\vspace{-0.3in}
\subsection{Hyperbolic User Embedding}

We learn user embeddings $\bm \theta^x_i$ in hyperbolic space, a Poincar\'{e} ball $\mathcal B^d$.
The basic idea is that, for each graph $\mathcal G^x$, $x \in \{s, t\}$, the proximity between users is preserved in the hyperbolic distance between user embeddings.


First, we conduct random walks to extract the proximity between users.
In a random walk, the neighborhood $\mathcal N^x_i$ of node $v^x_i$  is named as its ``context''.  
Intuitively, two nodes sharing more contexts are of higher proximity, and thus have similar embeddings \cite{perozzi2014deepwalk}.
In this case, each node is treated as a node for itself and a context for some other nodes.
Hence, to differentiate the user embedding $\bm \theta^x_i$ for itself, we introduce a context embedding ${\bm \theta^x_i}' \in \mathcal B^d$ for each node as well.

Then, we leverage the hyperbolic distance to preserve the proximity in $\mathcal B^d$. 
Specifically, we define the probability of having $v^x_j$ as a context of a given $v^x_i$ via the hyperbolic distance in Eq. (\ref{hdist}) as follows:
\vspace{-0.05in}
\begin{equation}
Pr(v^x_j|v^x_i)= {\sigma}[-d({\bm \theta^x_j}', \bm \theta^x_i)], 
\vspace{-0.03in}
\end{equation}
where $\sigma(x)=\frac{1}{\exp(-x)+1}$ is the sigmoid function.
For $\mathcal G^s$ and  $\mathcal G^t$, we minimize the negative log-likelihood as follows:
\vspace{-0.01in}
\begin{equation}
\mathcal O_{user}=  - \sum_{x \in \{s,t\}}\sum_{v^x_i \in \mathcal V^x}\sum_{v^x_j \in \mathcal N^x_i} \log Pr(v^x_j|v^x_i),
\label{obj1}
\vspace{-0.04in}
\end{equation}
\ie, we aim to construct the neighborhood of random walks via the hyperbolic distance between user embeddings.

\subsection{Hyperbolic Community Embedding}



With hyperbolic user embeddings above, we further learn community embeddings $\bm \mu^x_p$ in this Poincar\'{e} ball $\mathcal B^d$, where $p \in [1, C^x]$ and $C^x$ is the number of communities in $\mathcal G^x$.

Inspired by model-based clustering, 
we consider that 
user embeddings $\{\bm \theta^x_{\cdot}\}$ are drawn from the mixture of multivariate distribution $\{Pr(\cdot|\bm \psi^x_p)\}_{p=1}^{C^x}$  in hyperbolic space, where $\bm \psi_p^x$ is the distribution parameter to be described in details later. 
Each distribution $Pr(\cdot|\bm \psi^x_p)$ corresponds to a community $\mathcal C^p$, 
and the community embedding $\bm \mu^x_p$ is given as the location of $Pr(\cdot|\bm \psi^x_p)$ in the hyperbolic space.
Then, for all user embeddings $\{\bm \theta^x_{\cdot}\}$ in $\mathcal G^x$, we have the likelihood as follows:
\begin{equation}
\prod\nolimits_{i=1}^{N^x} \sum\nolimits_{p=1}^{C^x}  \bm Z_{ip} Pr(\bm \theta^x_i|\bm \psi^x_p),
\end{equation}
where $\bm Z_{ip}$ is the probability of user $v_i$ belonging to community $\mathcal C^p$, and thus we have $\bm Z_{ip} \in [0, 1]$ and $\sum\nolimits_{p=1}^{C^x}\bm Z_{ip}=1$.

We leverage (generalized) hyperbolic distribution to model communities in  hyperbolic space whose PDF is given as
\begin{equation}
\resizebox{1.02\hsize}{!}{$
\begin{aligned}
Pr_{\mathcal H}(\bm{\theta} ; \bm{\mu}, {\Delta}, \bm{\beta}, r, \omega )= &
\frac{e^{-\bm{\beta}^{T} \Delta^{-1}(\bm{\theta}-\bm{\mu})}}{(2 \pi)^{\frac{d}{2}}|{\Delta}|^{\frac{1}{2}}}
\left(\frac{\omega+\delta_{\bm \theta} }
{\omega+\bm \beta^T \Delta^{-1}\bm \beta}\right)
^{\frac{r-d/2}{2}} \\
&\frac{K_{r-\frac{d}{2}}\left(\sqrt{   (\omega+\bm \beta^T \Delta^{-1}\bm \beta) (\omega+\delta_{\bm \theta}) }\right)}{K_{r}(\omega)},
\end{aligned}
$}
 \label{pdf}
\end{equation}
where $ \delta_{\bm \theta}=(\bm{\theta}-\bm{\mu})^T {\Delta}^{-1}(\bm{\theta}-\bm{\mu})$. $\bm{\beta}\in \mathcal B^d $ and 
$\bm{\mu}\in \mathcal B^d $ are skewness and location vectors, respectively.  
$\omega$ is the concentration factor.
$\Delta \in \mathbb R^{d \times d}$ is the positive definite scatter matrix capturing Riemannian metric, 
and $|\Delta|$ is its determinant.
$K_{r}(\cdot)$ is the modified Bessel function of $(\cdot)$ with order $r$. 
Then, we have $\bm \psi_p^x=(\bm{\mu}_p^x, \Delta_p^x,\bm{\beta}_p^x, r_p^x, \omega_p^x)$ for each $x \in \{s, t\}$, and $\bm \theta_{i}^x \sim \sum\nolimits_{p=1}^{C^x} \bm Z^x_{ip} Pr_{\mathcal H}(\bm \theta_{i}^x|\bm \psi^x_p)$.
Thus, with user embeddings $\{\bm \theta^x_{i}\}$, we minimize the negative log-likelihood as follows:
\vspace{-0.05in}
\begin{equation}
\resizebox{0.91\hsize}{!}{$
 \mathcal O_{community}= - \sum\limits_{x \in \{s,t\}}\sum\limits_{v^x_i \in \mathcal V^x}\log\sum\limits_{p=1}^{C^x}\bm Z_{ip}^x Pr_{\mathcal H}(\bm \theta^x_i ; \bm \psi^x_p ),
$}
 \label{obj2}
 \vspace{-0.05in}
\end{equation}
so that we learn the community embedding and users' community membership simultaneously. 

\subsection{Hyperbolic Common Subspace}
We embed both users and communities for each network. Then, across $\mathcal G^s$ and $\mathcal G^t$, we construct a common hyperbolic subspace where we can jointly align users and communities. 

Specifically,  the common hyperbolic subspace is constructed via aligning embedding spaces of $\mathcal G^s$ and $\mathcal G^t$ on anchor users,
so that the user embedding of $v^t_k$ is transfered via anchor link $(v^s_i,v^t_k)$ to predict embeddings ${\bm \theta^s_{\cdot}}'$ in the neighborhood of its counterpart $v^s_i$.
We minimize the negative log-likelihood:
\begin{equation}
\resizebox{1.05\hsize}{!}{$
\begin{aligned}
 \mathcal O_{align}= &- \sum_{(v^s_i,v^t_k)\in \mathcal A} \left( \sum_{v^s_j\in \mathcal N^s_i} \log Pr(v^s_j|v^t_k) + \sum_{v^t_j\in \mathcal N^t_k} \log Pr(v^t_j|v^s_i) \right).
\end{aligned}
$}
\label{obj3}
\end{equation}
For each anchor user $(v^s_i,v^t_k) \in \mathcal A$, the first term is to predict embeddings in the neighborhood $\mathcal N^s_i$ of $ v^s_i$ in $\mathcal G^s$ using $\bm \theta^t_k$ of $\mathcal G^t$ and the second term is to predict embeddings in $\mathcal N^t_k$ using $\bm \theta^s_i$.
In this common subspace, communities as well as users are to be aligned via the hyperbolic distance.


\subsection{Objective Function}
Finally, we formulate the unified optimization of \textsc{Perfect} jointly aligning users and communities in the common hyperbolic subspace. We adopt the negative sampling \cite{mikolov2013distributed} to define the optimization objective, \ie, we replace the log term in $\mathcal O_{user}$ and $\mathcal O_{align}$ with the right hand side of the following equation and have $\mathcal O_{user}^{\text{NS}}$ and $\mathcal O_{align}^{\text{NS}}$. 
\begin{equation}
\vspace{-0.05in}
\begin{aligned}
 \log Pr(v_j^{t}|  v_i^{s} ) \  \propto & \ \log \sigma[-d({\bm \theta^{t}_j}', \bm \theta^{s}_i)] + \\
 & \sum\nolimits_{v_n \in \text{NS}^K_i} \mathbb E_{ v_n } \left[ \log \sigma[d({\bm \theta^{t}_n}', \bm \theta^{s}_i)]\right],
\end{aligned}
\label{neg}
\end{equation}
where the probability of negative sample $v_n$ being selected is as proposed in the study \cite{mikolov2013distributed}, and $\text{NS}^K_i$ is defined as a set of $K$ negative samples being selected randomly, $(v_n, v_i) \notin \mathcal E$.
Eq. (\ref{neg}) holds for identical superscripts of $v_i$ and $v_j$ as well.
Incorporating Eqs. (\ref{obj1}), (\ref{obj2}) and (\ref{obj3}), we finally obtain the objective function of \textsc{Perfect}:
\begin{equation}
\begin{aligned}
\min\limits_{ \bm \theta, \bm \theta', \bm \psi, \bm Z }  &\mathcal J_0 =  \mathcal O_{user}^{\text{NS}} + \alpha_1 \mathcal O_{community}^{} + \alpha_2 \mathcal O_{align}^{\text{NS}}\\
\text{s.t.} \quad  & \Delta^x_p \succeq 0,  p=1,2,..., C^x; \   \sum\nolimits_{p=1}^{C^x} \bm Z^x_{ip}=1,  \forall x \in \{s,t\},
\end{aligned}
\label{objall}
\vspace{-0.05in}
\end{equation}
where $\alpha_1$ and $\alpha_2$ are two nonnegative weight parameters. 

In the common subspace induced by optimizing Eq. (\ref{objall}), user embeddings will learn from community embeddings in both networks and, in turn, community embeddings learn from all user embeddings for \emph{the joint social network alignment}.




 \begin{algorithm}
          \caption{Alternating Riemannian Optimization}
          \LinesNumbered
            \KwIn{graph pair $ (\mathcal G^s, \mathcal G^t , \mathcal A )$, $\#$(community) $C^s$, $C^t$, embedding dimension $d$, $\#$(negative sample) $K$}
            \KwOut{user embeddings $\bm \theta^x$,  community embeddings  $\bm \mu^x$ and  membership $\bm Z^x$ for  $ x  \in  \{s,t\}$ 
            }
            Conduct random walks for each social network\;
            Initialize $\bm \theta^{x}$ and ${\bm \theta^{x}}'$ via $\min \mathcal O_{user}^{\text{NS}}$ for each $x \in \{s,t\}$\;
            \While{not converging} {
                \ForAll {social network} {
                              \ForAll {community $\mathcal C^x_p$ in each $ \mathcal G^x $}{
                                    Calculate the expectations in Eq. (\ref{estep})\;
                                    Obtain community membership $\bm Z^x$\;
                                    Update  $\{ \bm \psi_p^x \}_{p=1}^{C^x}$ via Eq. (\ref{mstep})\;
                              }
                              \ForAll {node $v^x_i$ in each $\mathcal G^x$} {
                                    Update $\bm \theta^x_i$ via Eqs. (\ref{user1})-(\ref{user3}) \;
                                    Update ${\bm \theta^x_i}'$ via  Eqs. (\ref{user4})-(\ref{user5})\;
                              }
                }
            }
\end{algorithm}
\vspace{-0.15in}
\section{Perfect: Optimization}\label{sec:opt}
To address the optimization in Eq. (\ref{objall}) , we first decompose this optimization problem into two subproblems, \ie, community embedding subproblem \wrt $(\bm \psi, \bm Z)$ and user embedding subproblem \wrt $(\bm \theta, \bm \theta')$. 
Then, we propose an alternating Riemannian optimization algorithm.
The main idea is that we alternatively optimize one subproblem while fixing the parameters of the other. 
The overall process is summarized in Algorithm 1.
In each iteration (Line $4$-$10$), community embeddings and user embeddings are mutually refined in the common hyperbolic subspace for the joint alignment. 
Every time we obtain the embeddings, the alignment is naturally revealed in the hyperbolic distance. 
We finally output the joint alignment results together until they cannot refine each other.
\subsection{Community Embedding Subproblem}
Given user embeddings, we update community embeddings 
for community alignment.
To facilitate the optimization, we first introduce auxiliary random variables of inverse Gaussian $W \sim \mathcal{I}(\omega, 1, r)$, $W \in \mathbb R$, 
and Gaussian $\bm g \sim  \mathcal N (\mathbf{0},  \Delta)$, $\bm g \in \mathbb R^d$. 
Then, a random variable $\bm \theta$  of the hyperbolic distribution with parameter $\bm \psi=(\bm{\mu},  \Delta, \bm{\beta}, r, \omega)$ is equal to the combination $\bm \mu+W \bm {\beta}+\sqrt{W} \bm g$ \cite{browne2015mixture}.  
Then, under the reformulation of this combination, we optimize $(\bm Z^x, \{ \bm \psi_p^x \}) $ via expectation-maximization algorithm.
Take $\mathcal G^s$ for instance.

In the expectation-step, we require following expectations:
\vspace{-0.03in}
\begin{flalign}
{z}^s_{ip} & =\mathbb{E}\left[\bm Z^s_{ip} | \bm{\theta}^s_{i}\right] = \frac{\bm Z^s_{p} Pr_{\mathcal H}\left(\bm{\theta}^s_{i} | \bm{\psi}^s_{p}\right)}{\sum_{p=1}^{C^s} \bm Z^s_{p} Pr_{\mathcal H}\left(\bm{\theta}^s_{i} | \bm{\psi}^s_{p}\right)},  \nonumber\\
a^s_{ip} &=\mathbb{E}\left[ W^s_{ip} | \bm{\theta}^s_{i}, \bm Z^s_{ip}=1\right]= \frac{K_{r^s_p+1}(\omega^s_{p}) }{K_{r^s_{p}}(\omega^s_{p})}-\frac{2 r^s_{p}}{\omega^s_{p} }, \nonumber \\
b^s_{ip}& =\mathbb{E}\left[ 1/W^s_{ip} | \bm{\theta}^s_{i}, \bm Z^s_{ip}=1\right]=\frac{K_{r^s_p+1}(\omega^s_{p})}{K_{r^s_{p}}(\omega^s_{p})},\nonumber\\
c^s_{ip} &=\mathbb{E}\left[ \log(W^s_{ip}) | \bm{\theta}^s_{i}, \bm Z^s_{ip}=1\right] =\frac{1}{K_{r^s_{p}}(\omega^s_{p})}\frac{\partial K_{r^s_{p}}(\omega^s_{p}) }{ \partial  r^s_{p}}, 
\label{estep}
\vspace{-0.1in}
\end{flalign}
where $\bm Z^s_{ip}$ is the membership of $i^{th}$ user  to $p^{th}$ community.
Then, we define: 
$n^s_{p}=\sum\nolimits_{i=1}^{N^s} {\bm Z}^s_{ip}$, 
$\overline{a}^s_{p}= \frac{1}{C^s}\sum\nolimits_{i=1}^{N^s} {\bm Z}^s_{ip} a^s_{ip}$, 
$\overline{b}^s_{p}= \frac{1}{C^s} \sum\nolimits_{i=1}^{N^s} {\bm Z}^s_{ip} b^s_{ip}$ 
and 
$\overline{c}^s_{p}= \frac{1}{C^s} \sum\nolimits_{i=1}^{N^s} {\bm Z}^s_{ip} c^s_{ip}$.

In the maximization-step, we derive the updating rules of hyperbolic distribution parameters 
as follows:
\vspace{-0.03in}
\begin{flalign}
\bm{\mu}^s_{p} &= \frac{\sum_{i=1}^{N^s} {\bm Z}^s_{ip} \bm{\theta}^s_{i}\left(\overline{a}^s_{p} b^s_{ip}-1\right)}{\sum_{i=1}^{N^s} {\bm Z}^s_{ip}\left(\overline{a}^s_{p} b^s_{ip}-1\right)},\nonumber  \\
\bm{\beta}^s_{p} &= \frac{\sum_{i=1}^{N^s} {\bm Z}^s_{ip} \bm{\theta}^s_{i}\left(\overline{b}^s_{p}-b^s_{ip}\right)}{\sum_{i=1}^{N^s} {\bm Z}^s_{ip}\left(\overline{a}^s_{p} b^s_{ip}-1\right)},\nonumber \\
\Delta^s_{p} &=   -{\bm{\beta}}^s_{p}\left(\overline{ \bm{\theta}}^s_{p}-{\bm{\mu}}^s_{p}\right)^T-\left(\overline{\bm{\theta}}^s_{p}-{\bm{\mu}}^s_{p}\right){{\bm{\beta}}^s_{p}}^T 
+\overline{a}^s_{p} {\bm{\beta}}^s_{p}   {  {\bm{\beta}}^s_{p} }^{T} \nonumber \\
 & \quad \ + \frac{1}{C^s} \sum\nolimits_{i=1}^{N^s} {\bm Z}^s_{ip} b^s_{ip}\left( \bm{\theta}^s_{i}-{\bm{\mu}}^s_{p}\right)\left( \bm{\theta}^s_{i}-{\bm{\mu}}^s_{p}\right)^{T}, 
 \label{mstep}
\end{flalign}
where 
$\overline{ \bm{\theta}}^s_{p}= \frac{1}{C^s}\sum\nolimits_{i=1}^{N^s}{\bm Z}^s_{ip}\bm{\theta}^s_{i}$.
We employ the strategy in the study \cite{browne2015mixture} to update $(r^s, \omega^s)$ and omit them due to the limit of space. 
The updating rules in network $\mathcal G^t$ are obtained by replacing the superscript $s$ with $t$. \\

\vspace{-0.07in}
\noindent \textbf{Correctness}: Note that, we estimate the hyperbolic distribution parameter  $\bm \psi=(\bm{\mu},  \Delta, \bm{\beta}, r, \omega)$ without considering the positive definite constraint of scatter matrices $\Delta$. 
Next, we prove that the positive definiteness is naturally guaranteed, \ie, \emph{Theorem 1 (Positive Definiteness)}, 
and thus show the correctness of the given updating rules.

\newtheorem*{lemma1}{Theorem 1 (Positive Definiteness)}
\begin{lemma1}
All scatter matrices of  $\Delta^{s}_p$ and $\Delta^{t}_p$ under updating rules in Eq. (\ref{mstep}) are positive definite.
\end{lemma1}
\begin{proof}
The proofs of $\Delta^s_p$ and $\Delta^t_p$ are the same, thus we use $\Delta^s_p$ for the elaboration. 
To prove the positive definiteness of $\Delta^s_p$, we first introduce an auxiliary matrix $\tilde{\Delta}^s_p$ defined as follows:
\begin{equation}
\tilde{\Delta}^s_p=\frac{1}{n_{p}} \sum_{i=1}^{N^s} z^s_{ip} b^s_{ip}\left(\bm{\theta}^s_{i}-{\boldsymbol{\mu}}_{p}^s-\frac{{\boldsymbol{\beta}}^s_{p}}{b^s_{ip}} \right)\left(\bm{\theta}^s_{i}-{\bm{\mu}}^s_{p}-\frac{{\boldsymbol{\beta}}^s_{p}}{c^s_{ip}} \right)^T.
\end{equation}


Note that, $a^s_{ip}=\mathbb{E}\left[ W^s_{ip} | \bm{\theta}^s_{i}, \bm Z^s_{ip}=1\right]$  and  $b^s_{ip}=\mathbb{E}[ 1/W^s_{ip} | \bm{\theta}^s_{i}, $   $\bm Z^s_{ip}=1]$. 
Based on Jensen's inequality, we have $1 / \mathbb{E}\left[W^s_{ip}\right] \leq \mathbb{E}\left[1 / W^s_{ip}\right]$ for all $i$, \ie, $1 / a^s_{ip} \leq b^s_{ip}$, and thus 
\begin{equation}
\overline{a}^s_{p}=\frac{1}{\bm Z^s_{p}} \sum_{i=1}^{N^s} \bm Z^s_{ip} a^s_{ip} \geq \frac{1}{\bm Z^s_{p}} \sum_{i=1}^{N^s} \frac{\bm Z^s_{ip}}{b^s_{ip}}.
\end{equation}
Finally, we have the following hold: 
\begin{equation}
\Delta^s_p = \tilde{\Delta}^s_p+ \left(\overline{a}^s_p -\frac{1}{\bm Z^s_p } \sum_{i=1}^{N^s} \frac{\bm Z^s_{ip}}{b^s_{ip}   }\right)    {\boldsymbol\beta}^s_p  {\boldsymbol\beta}_p^{s^T}.
\end{equation}
With the fact that   $\bm x \bm x^T$ is  positive definite  for any $\bm x \in \mathbb R^d$, we ensure the positive definiteness of $\Delta^s_p$ as well as $\Delta^t_p$.
\end{proof}

\subsection{User Embedding Subproblem}
With community embeddings fixed, we focus on updating user embeddings for user alignment.
These user embeddings live in the common Poincar\'{e} ball $\mathcal B^d$, a smooth unit manifold with Riemannian metric $\mathfrak{g}^{\mathcal B}(\bm \theta)$. 
In this manifold, the back-propagated gradient is Riemannian gradient and usual (Euclidean) gradient makes no sense as the operator of addition is not completely defined \cite{bonnabel2013stochastic}.

We optimize  $\{\bm \theta^{x}_i\}$ and $\{{\bm \theta_i^{x}}'\}$ via  exponential map with Riemannian gradient. 
Take $\mathcal G^s$ for instance. 
We first compute Riemannian gradient $\nabla_{\theta_{i}^s}^{R} \mathcal{J}$ of user  $\bm \theta^{s}_i$ to identify  optimizing direction in $T_{\bm \theta^{s}_i}\mathcal B^d$, 
and then leverage exponential map 
$
\bm \theta^s_{i} \leftarrow \exp _{\theta^s_{i}}\left(-\rho \nabla_{\theta^s_{i}}^{R} \mathcal{J}\right)
$
to move $\bm \theta^s_{i}$ along the mapped geodesic in the common Poincar\'{e} ball $\mathcal B^d$ with a step size $\rho$ \cite{bonnabel2013stochastic}. 
Fortunately, owing to the conformality of  $\mathcal B^d$, Riemannian gradient  $\nabla_{\theta }^{R} \mathcal{J}$ is obtained by rescaling the Euclidean gradient $\nabla_{\theta }^{E} \mathcal{J}$, \ie,  
$
\nabla_{\theta }^{R} \mathcal{J}=\left(\frac{1}{\lambda_{\theta }}\right)^{2} \nabla_{\theta }^{E} \mathcal{J}$.
The exponential map $\exp_{\theta}(\bm a)$ at $\bm \theta \in \mathcal B^d$ is defined in the following fraction \cite{ganea2018hyperbolic}:
\begin{equation}
\resizebox{1.02\hsize}{!}{$
\begin{aligned}
\frac{\lambda_{\theta}\theta \left(\cosh \left(\lambda_{\theta}\|\bm a\|\right)+\left\langle\theta, \frac{\bm a}{\|\bm a\|}\right\rangle \sinh \left(\lambda_{\theta}\|\bm a\|\right)\right) + \frac{\bm a}{\|\bm a\|}\sinh \left(\lambda_{\theta}\|\bm a\|\right)}
{1+\left(\lambda_{\theta}-1\right) \cosh \left(\lambda_{\theta}\|\bm a\|\right)+\lambda_{\theta} \left\langle \bm \theta, \frac{\bm a}{\|\bm a\|}\right\rangle \sinh \left(\lambda_{\theta}\|\bm a\|\right)}.
\end{aligned}
$}
\end{equation}



The remaining challenge lies in the challenge of obtaining Euclidean gradients
owing to summation within logarithm. 
To address this challenge,  we optimize an upper bound objective function $\mathcal J_1=  \mathcal O_{user}^{\text{NS}} + \alpha_1 \mathcal O_{community}^{\text{UP}} + \alpha_2 \mathcal O_{align}^{\text{NS}}$ instead of $\mathcal J_0$.
Specifically, we replace $\mathcal O_{community}$ with its upper bound $\mathcal O^{\text{UP}}_{community}$ 
defined as follows:
\vspace{-0.1in}
\begin{equation}
 \mathcal O^{\text{UP}}_{community}= - \sum_{x \in \{s,t\}}\sum_{v^x_i \in \mathcal V^x}\sum_{p=1}^{C^x}\bm Z^x_{ip}\log Pr_{\mathcal H}(\bm{\theta}^x_p ; \bm{\psi}^x_p ).
 \end{equation}
It is easy to be verified via the log-concavity:
$$\log \sum\nolimits_{p=1}^{C^x} \bm Z_{ip} Pr_{\mathcal H}(\bm{\theta}^x_p ; \bm{\psi}^x_p ) \ge \sum\nolimits_{p=1}^{C^x} \bm Z_{ip}  \log  Pr_{\mathcal H}(\bm{\theta}^x_p ; \bm{\psi}^x_p ).$$
Thus, we have the partial derivative of $\mathcal J_1$ \wrt $\bm \theta^s_i$, \ie, 
  $\frac{\partial }{\partial \bm \theta^s_i} \mathcal J_1= \frac{\partial \mathcal O_{user}^{\text{NS}}  }{\partial \bm \theta^s_i} +\alpha_1 \frac{\partial \mathcal O_{community}^{\text{UP}}}{\partial \bm \theta^s_i}  +\alpha_2 \mathbb I_{(v^s_i,v^t_k)\in \mathcal A} \frac{\partial  \mathcal O_{align}^{\text{NS}}}{\partial \bm \theta^s_i} $, 
\resizebox{1.03\linewidth}{!}{
  \begin{minipage}{1.19\linewidth}
\begin{flalign}
\frac{\partial \mathcal O_{user}^{\text{NS}}  }{\partial \bm \theta^s_i} = 
  & - \sum\limits_{v^s_j\in \mathcal N^s_i } \left(
  \sigma [d({\bm \theta^s_j}', \bm \theta^s_i)]  \frac{\partial d({\bm \theta^s_j}', \bm \theta^s_i)}{\partial \bm \theta^s_i}\  -  \right. \nonumber \\
  &  \quad \quad \quad \ \  \left.  \sum\limits_{v_n \in \text{NS}^K_i} \mathbb E_{ v_n }\left[ \sigma [-d({\bm \theta^s_n}', \bm \theta^s_i)] \frac{\partial d({\bm \theta^s_n}', \bm \theta^s_i)}{\partial \bm \theta^s_i} \right]  \right), \label{user1}
\\
\frac{\partial \mathcal O_{community}^{\text{UP}}}{\partial \bm \theta^s_i} = 
 & - \sum\limits_{p = 1}^{C^s} {{\bm \pi^s_{p}}\left((\Delta^s_p)^{-1}(\bm \beta_p^s  +     \frac{\zeta_p^s  \tilde{\bm \theta^s_i}}{  {\tilde \delta }_{\theta^s_i}   } ) + \frac{\partial  \log K_{\zeta^s_p}(\sqrt{\nu^s_p {\tilde \delta }_{\theta^s_i}}  ) }{\partial  \bm \theta^s_i}
  \right)}, \label{user2}
  \\
\frac{\partial  \mathcal O_{align}^{\text{NS}}}{\partial \bm \theta^s_i}  = 
  & - \sum\limits_{v^t_j\in \mathcal N^t_k }
  \left( \sigma [d({\bm \theta^t_j}', \bm \theta^s_i)]  \frac{\partial d({\bm \theta^t_j}', \bm \theta^s_i)}{\partial \bm \theta^s_i} \ - \right.  \nonumber \\
 &  \quad \quad \quad \ \   \left. \sum\limits_{v_n \in \text{NS}^K_i} \mathbb E_{ v_n } \left[\sigma [-d({\bm \theta^s_n}', \bm \theta^s_i)] \frac{\partial d({\bm \theta^t_n}', \bm \theta^s_i)}{\partial \bm \theta^s_i} \right] \right), \label{user3}
\end{flalign}
\vspace{0.01in}
\end{minipage}
}
where $\nu^s_p= \omega^s_p +{\bm \beta^s_p}^T(\Delta^s_p)^{-1}{\bm \beta^s_p} $, 
$\zeta^s_p = r^s_p -d/2$, 
${\tilde { \bm \theta}}^s_i=  \bm \theta^s_i- \bm \mu^s_p$, 
${\tilde \delta }_{\theta^s_i}={ \delta }_{\theta^s_i}+\omega^s_p$
and $\sigma(x)$ is the standard sigmoid function.
$\mathbb I_{(\cdot)}$ returns $1$ iff the condition $(\cdot)$ is true; otherwise, $\mathbb I_{(\cdot)}=0$. 
$\frac{\partial d ({\bm \theta_1}, \bm \theta_2)}{\partial \bm \theta_1} $ is given in Eq. (\ref{pdofd}).
Utilizing the fact $\frac{\partial }{\partial x}K_{r}(x)=-\frac{r}{x} K_{r}(x)-K_{r-1}(x)$, we give the partial derivative  of $  \log K_{\zeta^s_p}(\sqrt{\nu^s_p {\tilde \delta }_{\theta^s_i} }  )$ \wrt $\bm \theta^s_i$ as follows:
\begin{equation}
\resizebox{1.03\hsize}{!}{$
\begin{aligned}
\frac{\partial \log K_{\zeta^s_p}(\sqrt{\nu^s_p {\tilde \delta }_{\theta^s_i} }  )}{\partial  \bm \theta^s_i}  =  
- & \left( \frac{\zeta^s_p }{   {\tilde \delta }_{\theta^s_i}  } 
+ \sqrt {   \frac{\nu^s_p }{   {\tilde \delta }_{\theta^s_i} } } 
\frac{K_{\zeta_p}(   \sqrt{   \nu^s_p {  {\tilde \delta }_{\theta^s_i}  }  } )   
}{K_{\zeta_p-1}(   \sqrt{  \nu^s_p {   {\tilde \delta }_{\theta^s_i}   }   }  )
}\right)   
\left(\Delta^s_p\right) ^{ - 1}{\tilde { \bm \theta}}^s_i.
\nonumber
\end{aligned}
$}
\end{equation}

Similarly, for ${\bm \theta^s_j}'$, we have $\frac{\partial \mathcal J_1}{\partial {\bm \theta^s_j}'}= \frac{\partial \mathcal O^{\text{NS}}_{user}}{\partial {\bm \theta^s_j}'}+ \alpha_1 \frac{\partial \mathcal O^{\text{NS}}_{align}}{\partial {\bm \theta^s_j}' }$,
\resizebox{1.03\linewidth}{!}{
  \begin{minipage}{1.18\linewidth}
\begin{flalign}
\frac{\partial \mathcal O^{\text{NS}}_{user}}{\partial { \bm \theta^s_j}'} = 
  & - \sum\limits_{v^s_i\in \mathcal V^s}
  \left( \mathbb I_{v^s_j \in \mathcal N^s_i} \sigma [d({\bm \theta^s_j}', \bm \theta^s_i)]  \frac{\partial d({\bm \theta^s_j}', \bm \theta^s_i)}{\partial {\bm \theta^s_j}'} - \right. \nonumber \\
  & \quad \quad \quad  \left. \sum\limits_{v_n \in \text{NS}^K_i} \mathbb E_{ v_n }
  \left[ \mathbb I_{v^s_j=v^s_n}  \sigma [-d({\bm \theta^s_n}', \bm \theta^s_i)]  \frac{\partial d({\bm \theta^s_n}', \bm \theta^s_i)}{\partial {\bm \theta^s_n}'} \right] \right), \label{user4}
\\
\frac{\partial \mathcal O^{\text{NS}}_{align}}{\partial { \bm \theta^s_j}'} = 
  & - \sum\limits_{(v^s_i,v^t_k)  \in \mathcal A}
\left( \mathbb I_{v^s_j \in \mathcal N^s_i} \sigma [d({\bm \theta^s_j}', \bm \theta^t_k)] \frac{\partial d({\bm \theta^s_j}', \bm \theta^t_k)}{\partial {\bm \theta^s_j}'} - \right. \nonumber \\
  & \quad \quad \quad \quad \  \left. \sum\limits_{v_n \in \text{NS}^K_i} \mathbb E_{ v_n }
 \left[ \mathbb I_{v^s_j=v^s_n}  \sigma [-d({\bm \theta^s_n}', \bm \theta^t_k)]  \frac{\partial d({\bm \theta^s_n}', \bm \theta^t_k)}{\partial {\bm \theta^s_n}'}\right] \right). \label{user5}
\end{flalign}
\end{minipage}
}

\noindent Note that, $\mathbb I_{(\cdot)}$  is defined above. The updating rules in network $\mathcal G^t$ are obtained by swapping the superscript $s$ and $t$.

\noindent \textbf{Remarks}: In the common Poincar\'{e} ball, Eq. (\ref{user2}) encourages users to join in their corresponding communities, the alignment of which is inferred by user embeddings within the community. 
Hence,  user embeddings and community embeddings are mutually refined  for the joint social network alignment.

\noindent \textbf{Computational Complexity:} 
We analyze the computational complexity of Algorithm 1. 
First, we generate random walks (Line 1), costing $O(Nhl)$, $N:=\max\{N^s, N^t\}$, where $h$ is the number of walks per node and $l$ is the walk length.
Then, we solve two subproblems alternatively, which tends to be converged in a few iterations.
Specifically, updating the parameters of community embedding subproblem (Line 5-7) costs $O(C^2d^2)$, where $C:=\max\{C^s, C^t\}$,
and $O(Nhl(\epsilon Kd+Cd^2))$ for user embedding subproblem (Line 8-10), where $\epsilon$ is the size of the node neighborhood.
To sum up, the overall complexity of \textsc{Perfect} is linear to the number of users $N$.

\vspace{-0.1in}
\section{Experiments} \label{sec:exp}
\vspace{-0.1in}
In \textsc{Perfect}, we jointly align communities and users in the hyperbolic space. 
Thus, we evaluate the performance of \textsc{Perfect} with baseline methods on both community alignment and user alignment in the experiments, whose results are reported in Sections \ref{sec:exp-com} and \ref{sec:exp-user}, respectively.
Datasets are introduced in Section \ref{sec:why}.
We repeat each experiment $10$ times and report the mean with $95\%$ confidence interval. 

\vspace{-0.12in}
\subsection{Experiments on Community Alignment}\label{sec:exp-com}
\vspace{-0.05in}
In this part, we  provide the  performance of \textsc{Perfect} compared against comparison methods on community alignment.
\subsubsection{Comparison Methods} To the best of our knowledge, the only method \cite{chen2017community} considering community alignment is designed for attributed networks specifically, which will not work for normal networks without node attributes and thus is not comparable in our experiments.
Specifically, the baseline methods compared here include both classic network embedding methods, \eg, DeepWalk and LINE and the latest ones with considerations of community structures, \ie, CommGAN:
\begin{itemize}
\item \emph{CommGAN} \cite{jia2019communitygan}: It is a recent method using a minimax game to update node embeddings for community discovery. For each network, we perform CommGAN and calculate mean embedding in each community. Community alignment is obtained by calculating Euclidean distance. 
\item \emph{DeepWalk} \cite{perozzi2014deepwalk}: For each network, we perform DeepWalk to obtain node embeddings and then employ K-Means to discover communities. We align communities by Euclidean distance between mean embeddings.
\item \emph{LINE} \cite{tang2015line}: We embed each network via LINE and use K-Means to discover communities, which will be aligned by Euclidean distance between mean embeddings.
\end{itemize}
To further evaluate \textsc{Perfect}, we design several variants of it:
\begin{itemize}
\item \emph{\textsc{Perfect}$-$}: To demonstrate the superiority of unified optimization, we design a two-step method, \ie, we learn the common subspace and community embeddings separately. Specifically, we first learn the common subspace  via optimizing $\mathcal O^{\text{NS}}_{user}+\mathcal O^{\text{NS}}_{align}$ and then obtain community embeddings via optimizing $\mathcal O^{\text{UP}}_{community}$. 
\item \emph{EucAlign}: To demonstrate the superiority of  hyperbolic space, we consider the corresponding Euclidean version of \textsc{Perfect}.  We only replace the hyperbolic distance $d(\bm x, \bm y)$ in \textsc{Perfect} with Euclidean distance.
\item \emph{EucAlign}$-$: It is the Euclidean version of \textsc{Perfect}$-$ where the Euclidean distance is used.
\end{itemize}

\vspace{-0.1in}
\subsubsection{Evaluation Metric} 
\begin{itemize}
\item \emph{$Accuracy=  \frac {1}{N_C} \sum_{i=1}^{N_C}  success_\tau(\mathcal C_p^s, \mathcal C_q^t) $}, where $N_C$ is the number of groundtruth anchor communities.
 $ success_\tau(\mathcal C_p^s, \mathcal C_q^t)$ returns $1$ iff groundtruth anchor community $(\mathcal C_p^s, \mathcal C_q^t)$ is successfully identified \wrt the threshold $\tau$; otherwise, it returns $0$. 
\item \emph{$Quality = 2 \mathbb E_i \{\sigma[-dist(\bm \mu^s_p, \bm \mu^t_q)] \}$}, where the sigmoid $\sigma(\cdot)$ is for normalization.
$ \bm \mu^s_p$ and $\bm \mu^t_q$ are community embeddings of groundtruth anchor community $\mathcal C_p^s$ and $\mathcal C_q^t$, respectively.
We use $Quality$ to evaluate community  alignment in the embedding space, and $dist(\cdot, \cdot)$ is the distance function of corresponding comparison methods.
\end{itemize}

\begin{figure} 
\centering 
\subfigure[DBLP-AMiner: Accuracy]{
\includegraphics[width=0.49\linewidth]{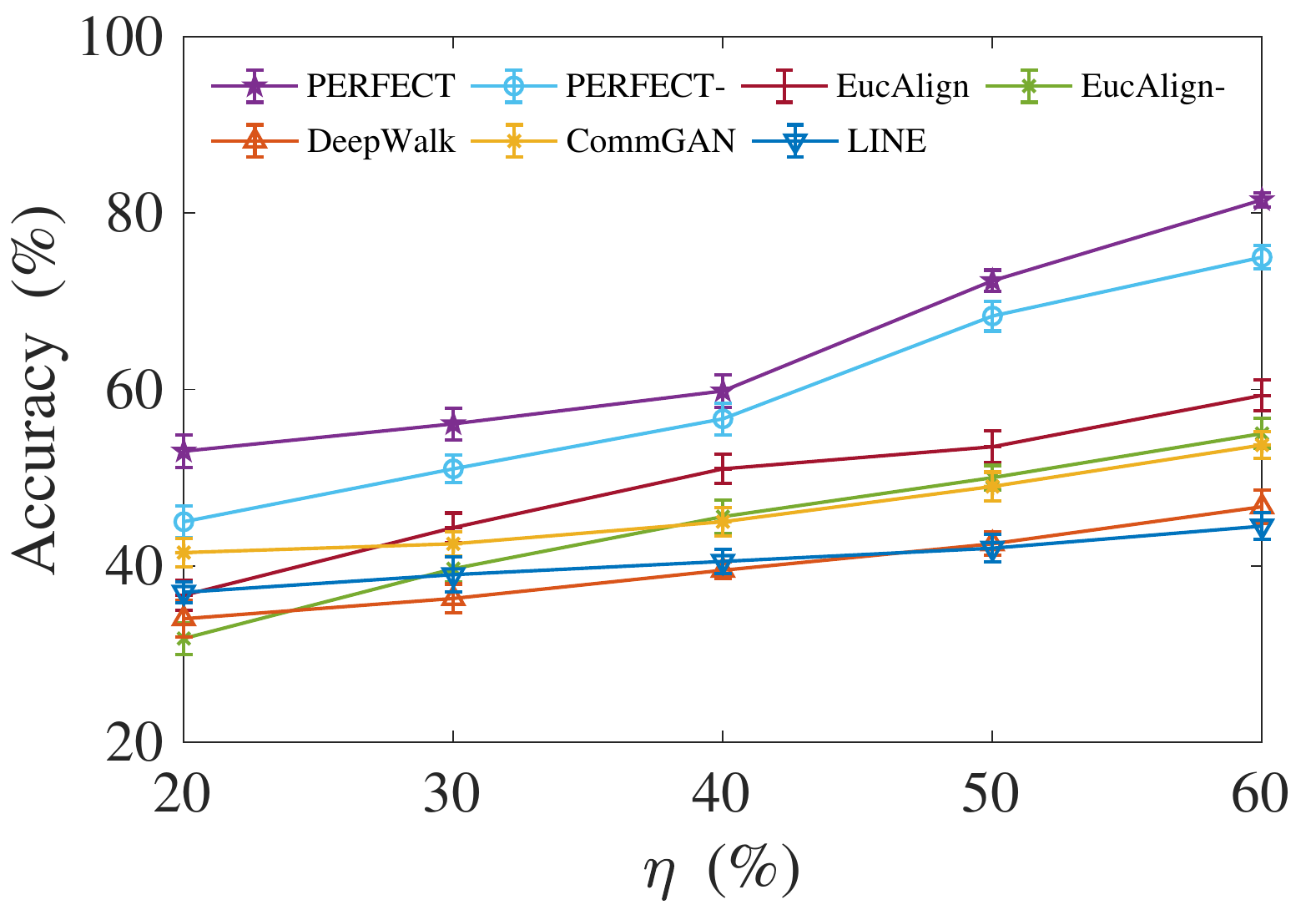}}
\hspace{-0.036\linewidth}
\subfigure[Twitter-Quora: Accuracy]{
\includegraphics[width=0.49\linewidth]{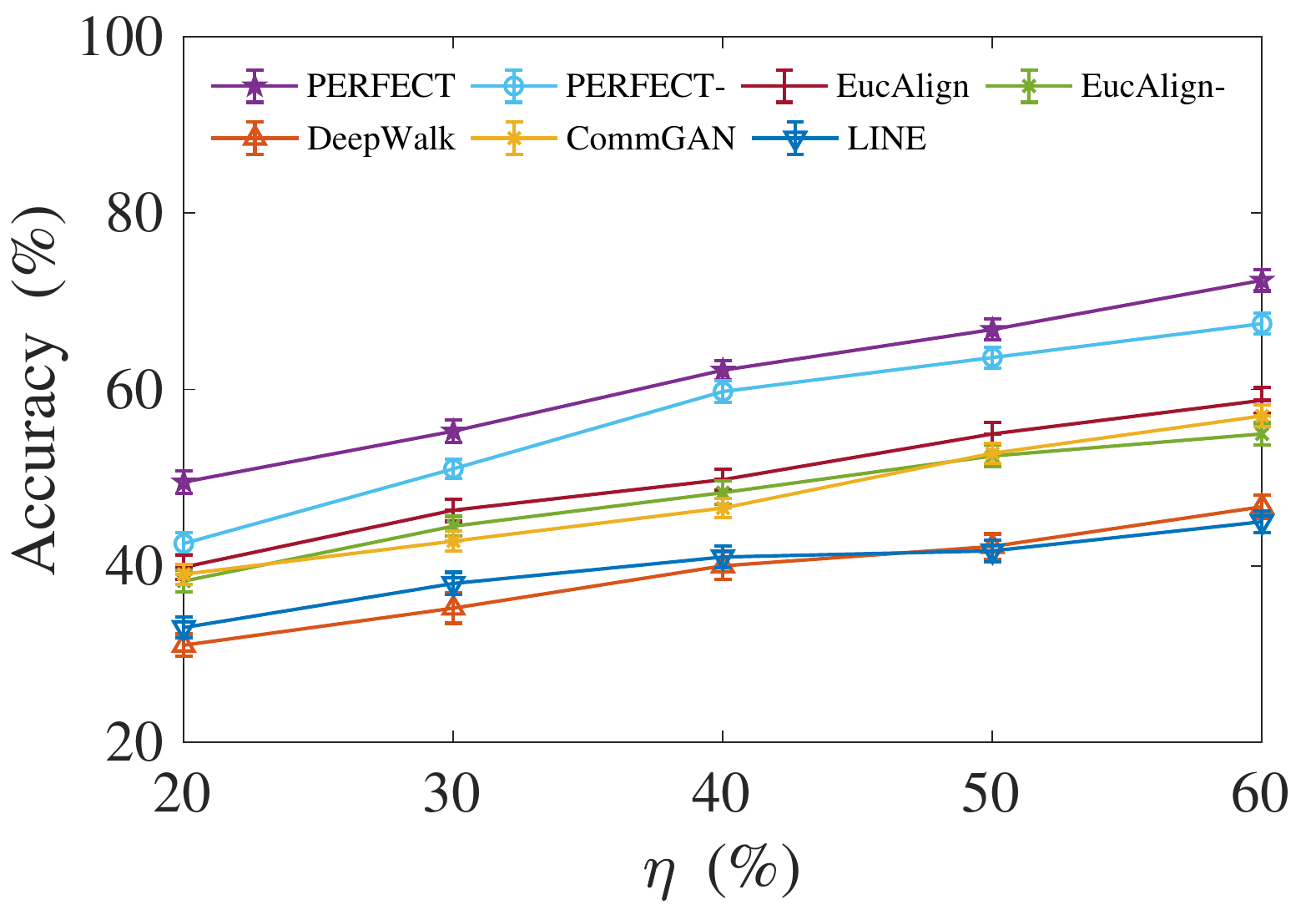}}
\vfill
\subfigure[DBLP-AMiner: Quality]{
\includegraphics[width=0.49\linewidth]{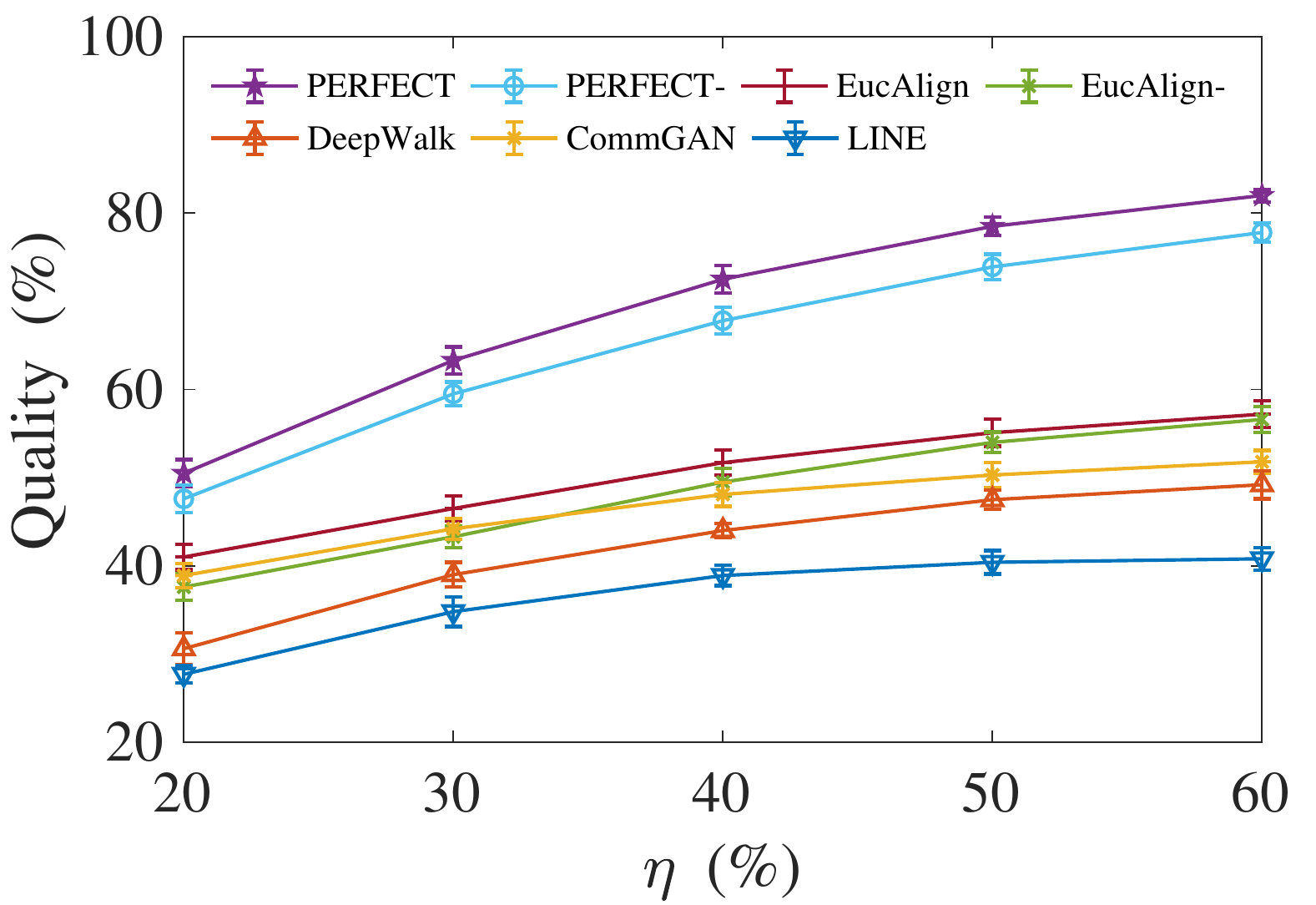}}
\hspace{-0.036\linewidth}
\subfigure[Twitter-Quora: Quality]{
\includegraphics[width=0.49\linewidth]{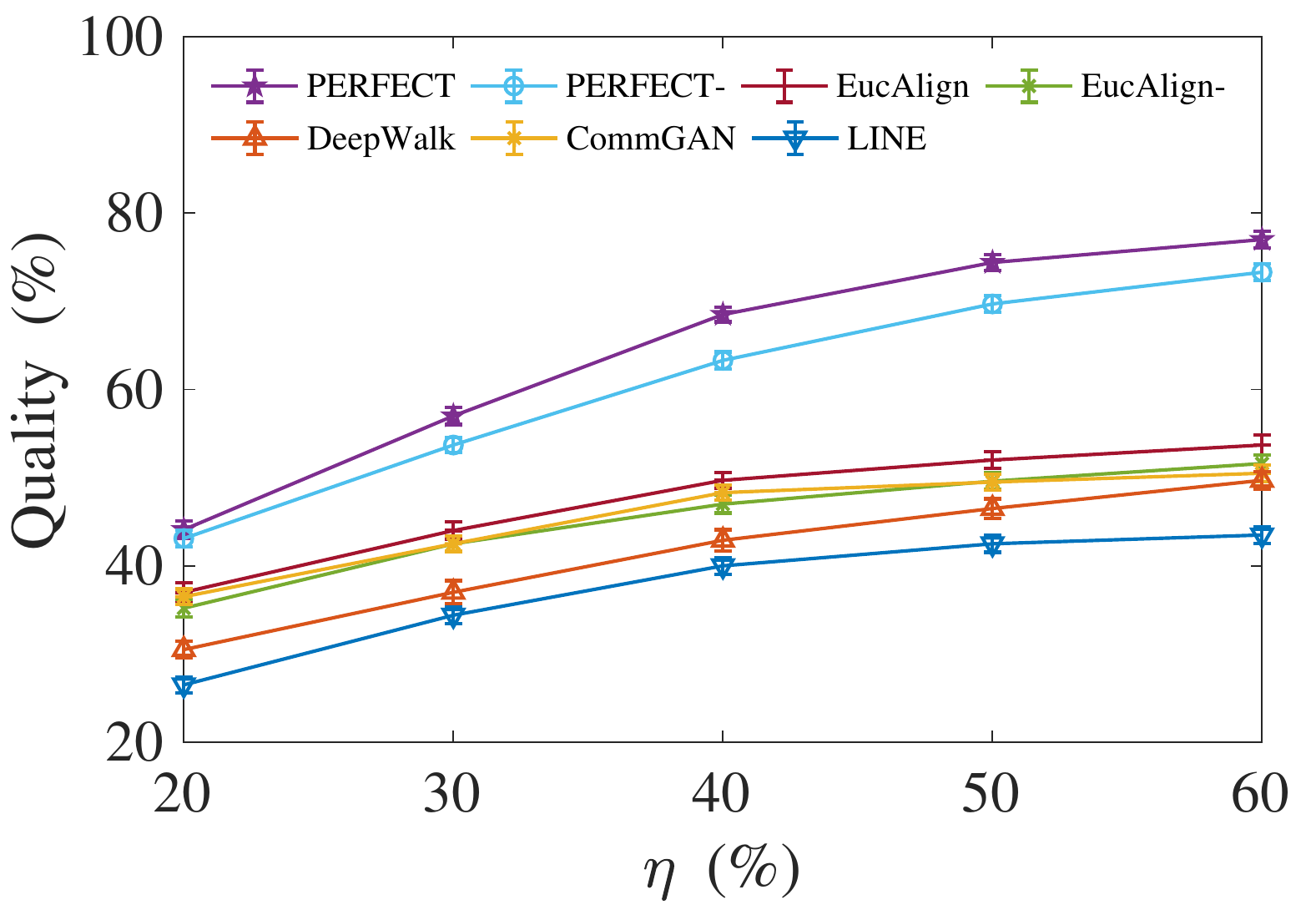}}
\vspace{-0.1in}
\caption{Community alignment under different overlaps}
\vspace{-0.1in}
\label{comAlignEta}
\end{figure}
\begin{figure}
\subfigure[DBLP-AMiner: Accuracy]{
\includegraphics[width=0.49\linewidth]{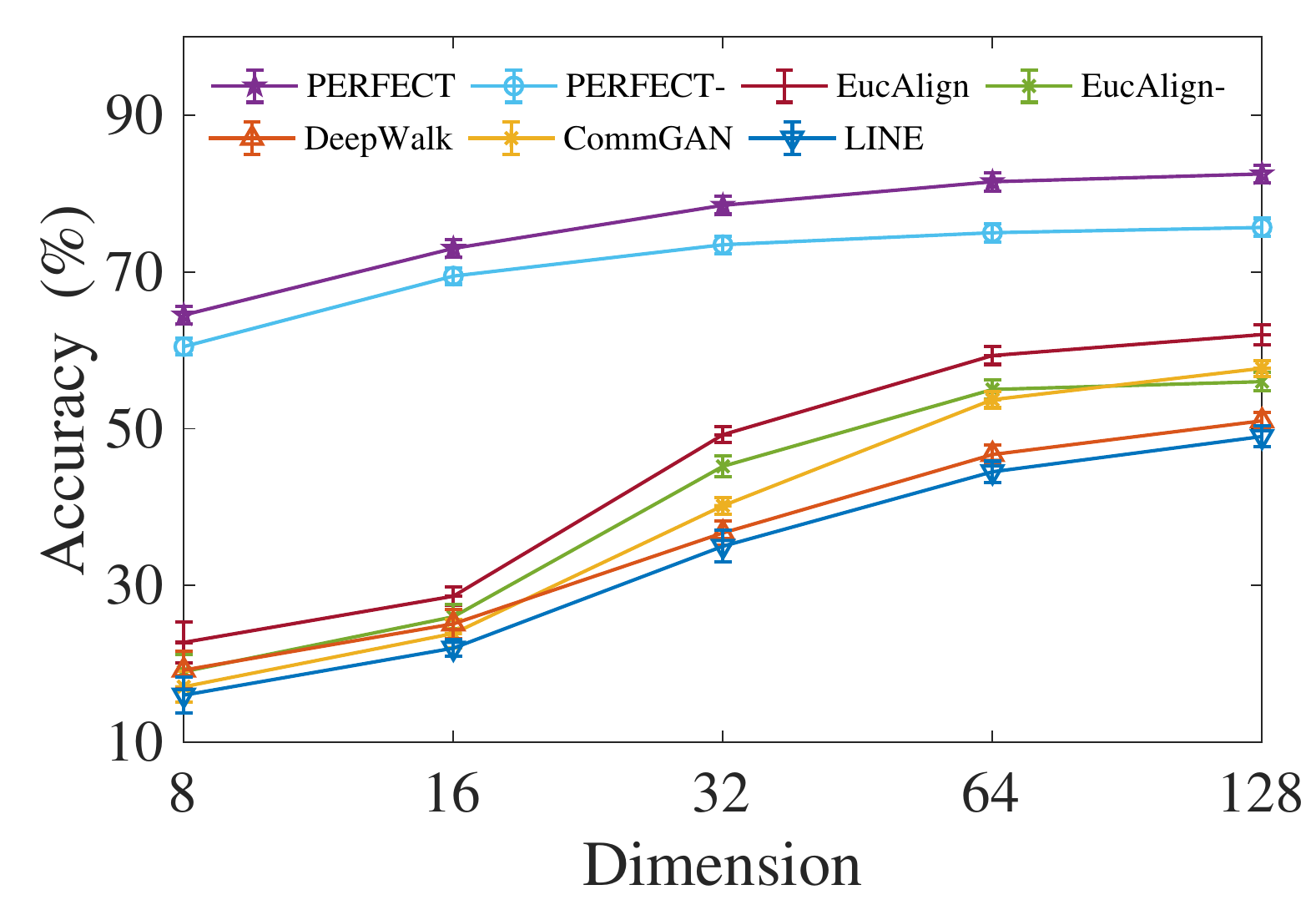}}
\hspace{-0.036\linewidth}
\subfigure[Twitter-Quora: Accuracy]{
\includegraphics[width=0.49\linewidth]{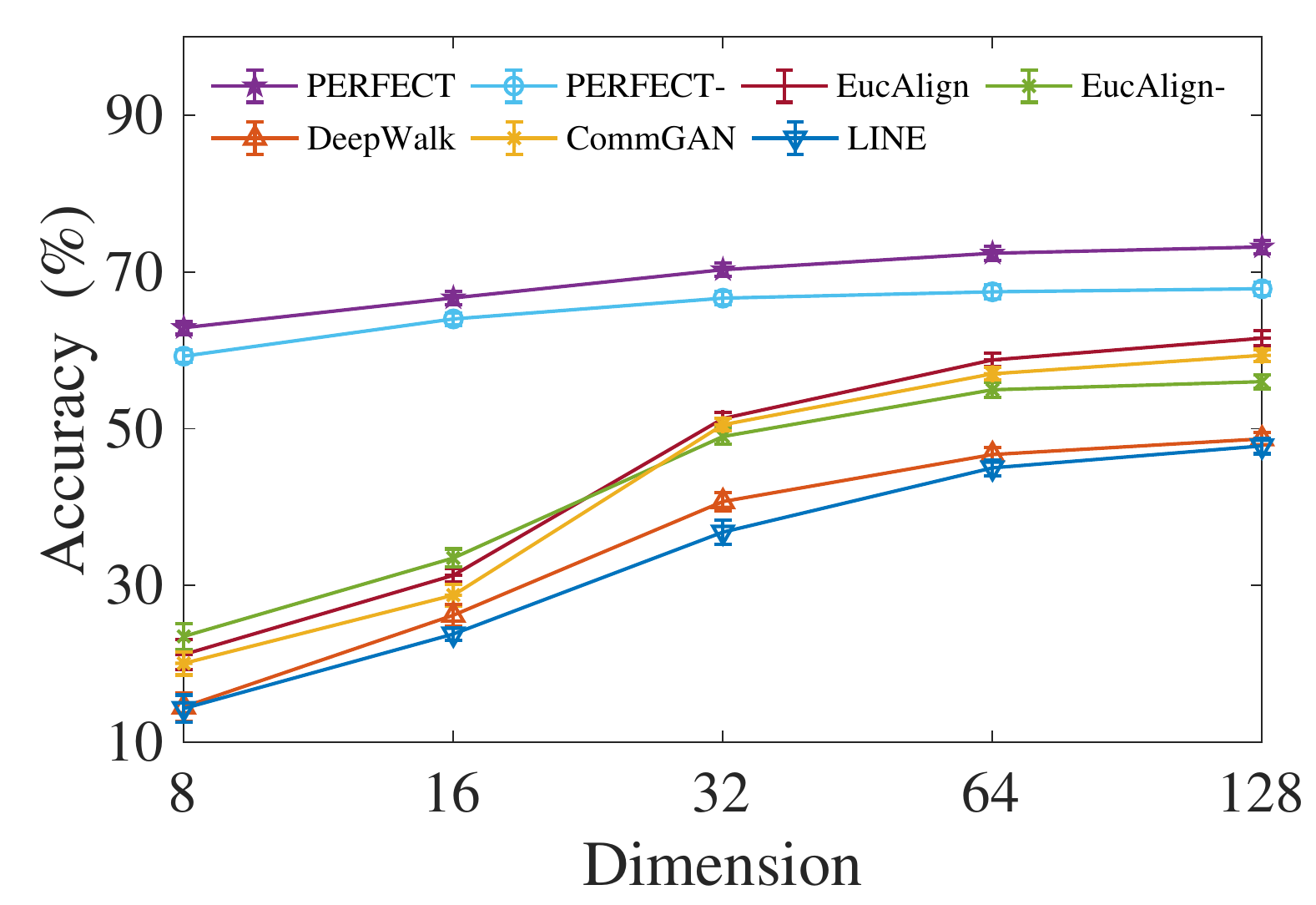}}
\vfill
\subfigure[DBLP-AMiner: Quality]{
\includegraphics[width=0.49\linewidth]{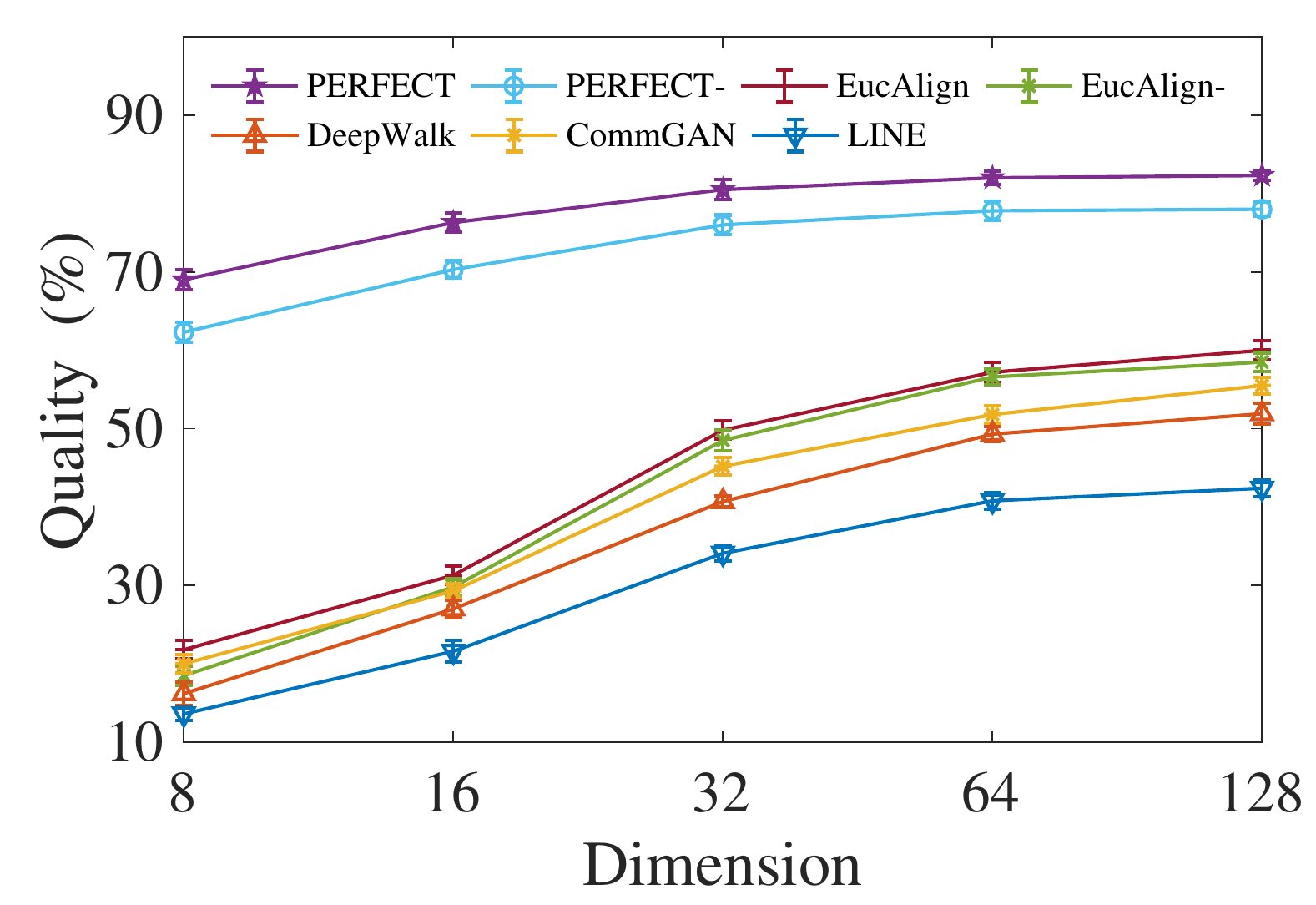}}
\hspace{-0.036\linewidth}
\subfigure[Twitter-Quora: Quality]{
\includegraphics[width=0.49\linewidth]{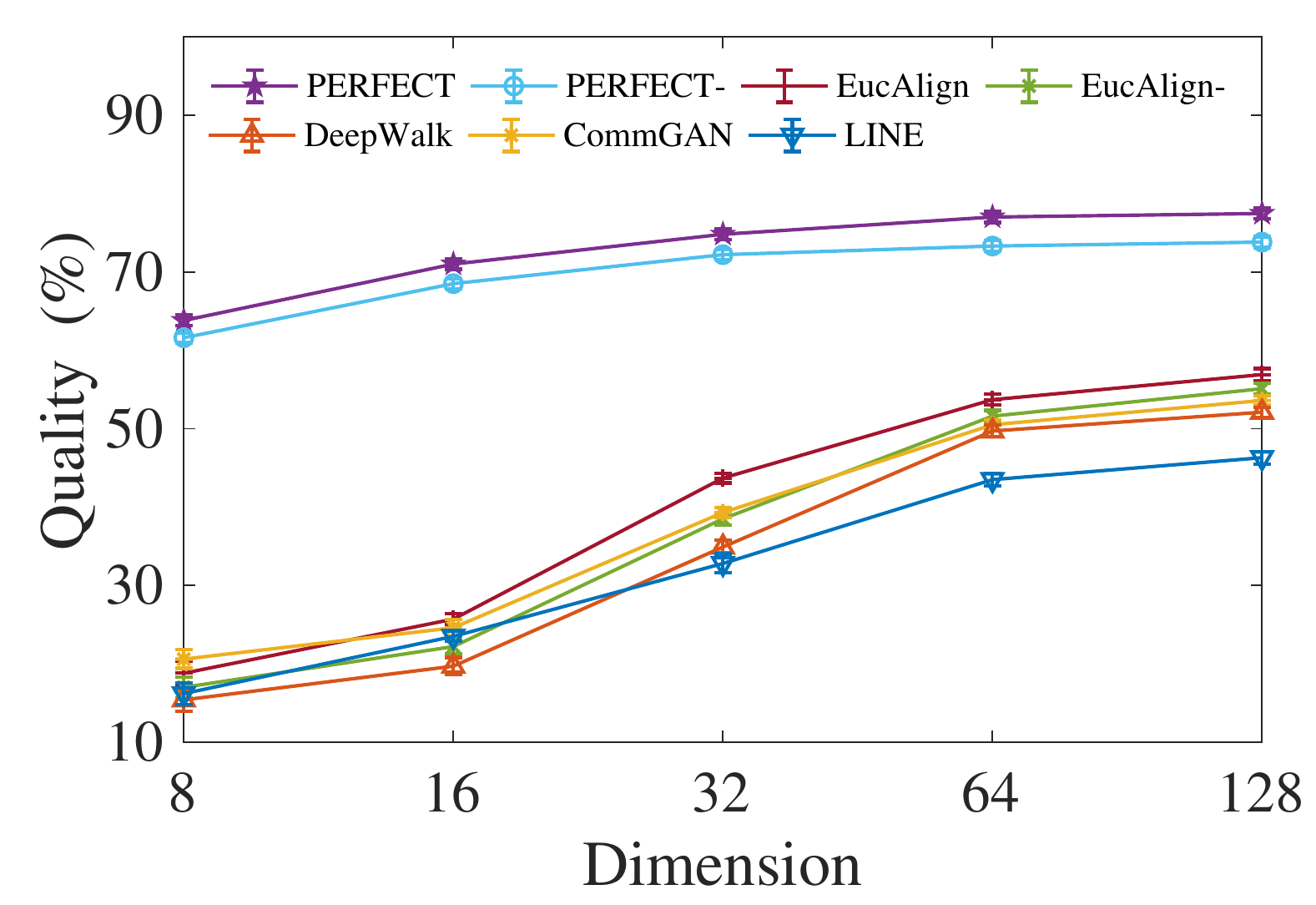}}
\vspace{-0.1in}
\caption{Community alignment under different dimensions}
\vspace{-0.1in}
\label{comAlignDi}
\end{figure}
\begin{figure}
\subfigure[DBLP-AMiner: Accuracy]{
\includegraphics[width=0.49\linewidth]{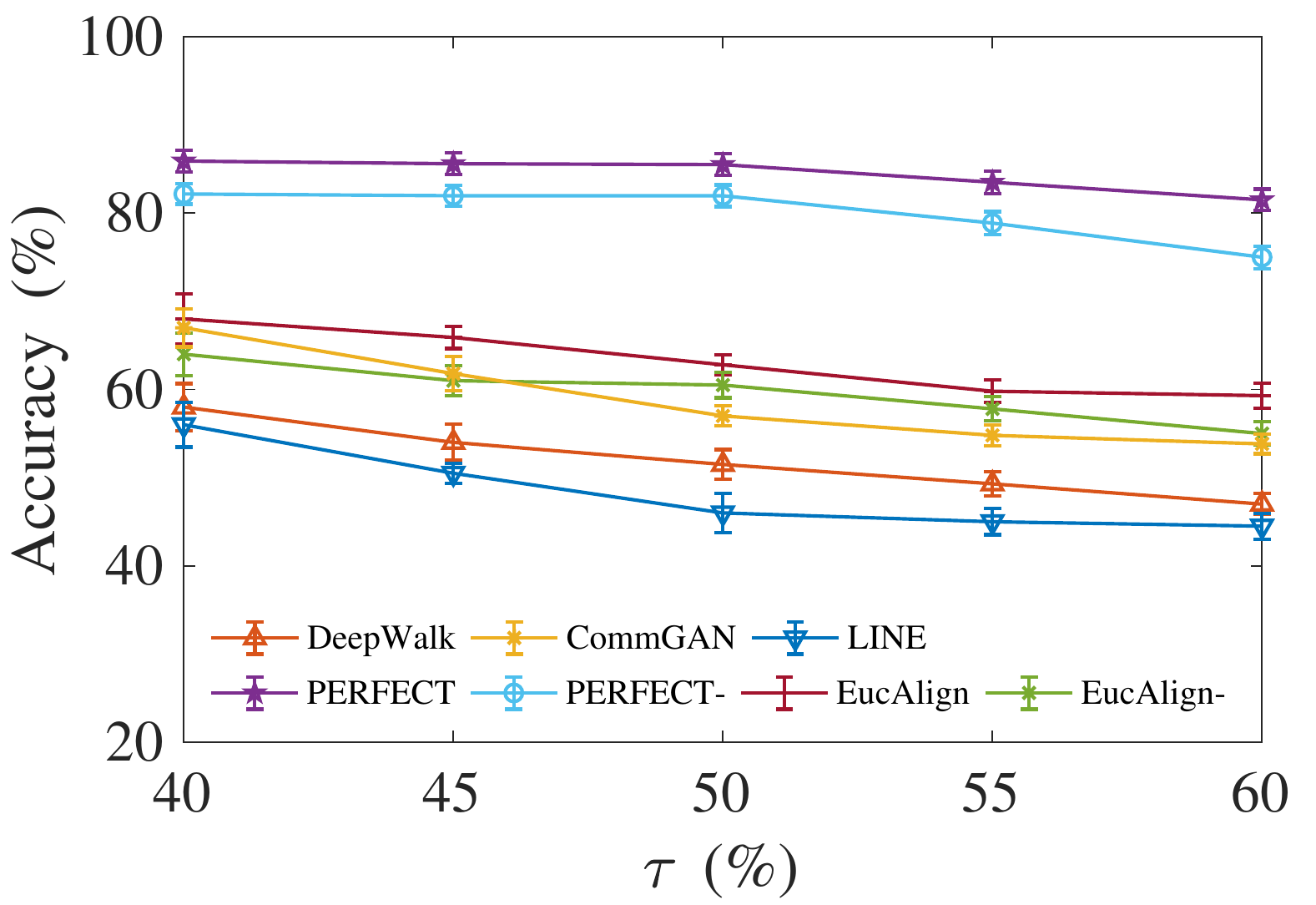}}
\hspace{-0.036\linewidth}
\subfigure[Twitter-Quora: Accuracy]{
\includegraphics[width=0.49\linewidth]{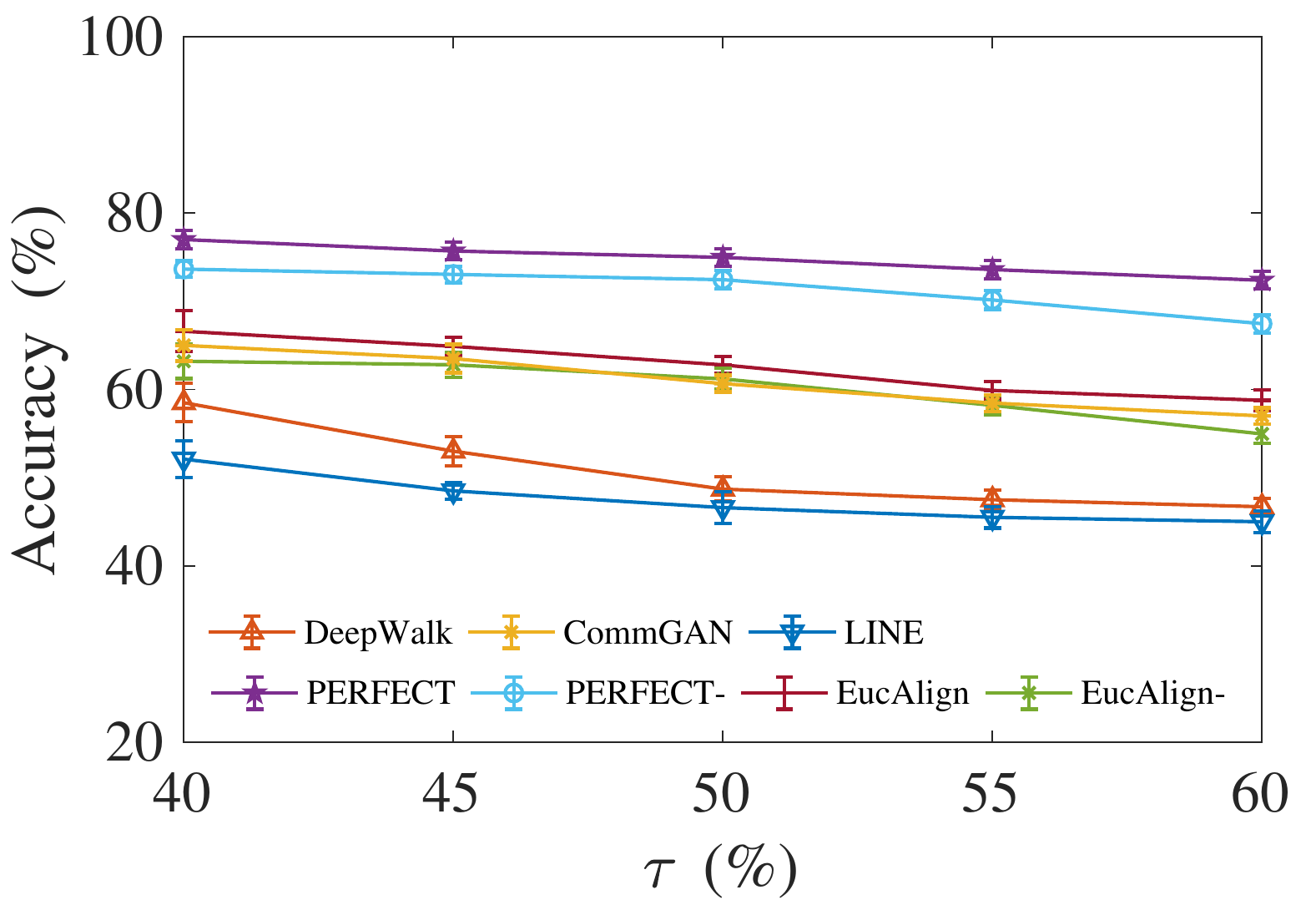}}
\vspace{-0.15in}
\caption{Parameter sensitivity of the threshold $\tau$}
\vspace{-0.15in}
\label{comAlignTau}
\end{figure}

\begin{table*}[htbp]
  \vspace{-0.03 in}
  \caption{The Precision of user alignment on DBLP-AMiner and Twitter-Quora datasets (\%)}
  \vspace{-0.07 in}
    \scriptsize
    \centering
    \resizebox{1\textwidth}{!}{
      \begin{tabular}{p{1.2cm}<{\centering}| p{1.5cm}<{\centering} | p{1.7cm}<{\centering}   p{1.7cm}<{\centering} p{1.7cm}<{\centering}  p{1.7cm}<{\centering} p{1.7cm}<{\centering} }
        \hline
        {\textbf{Dataset}} &{\textbf{Method}} & {$\textbf{k=10}$}&  {$\textbf{k=15}$}& {$\textbf{k=20}$}& {$\textbf{k=25}$}& {$\textbf{k=30}$}\\
           \hline
            \multirow{6}{*}{\tabincell{c}{DBLP\\$\&$\\AMiner} }
             & PALE&   $ 24.57\pm2.23 $   &   $ 31.36\pm1.31$  &   $46.22\pm0.98$  & $ 52.37\pm1.33$  &   $ 57.81\pm1.51$  \\  
              &IONE &   $ 29.05\pm1.13$   &   $ 35.21\pm1.45$  &   $48.63\pm1.58$   &   $ 53.45\pm1.13 $ &   $59.59 \pm1.41$  \\
              & SNNA&  $32.46\pm1.78$   &   $ 37.53\pm1.85$  &   $ 43.26\pm2.17$   &  $57.42\pm2.04$  &   $63.09 \pm1.82$  \\
          & MOANA&   $26.82\pm1.37$   &  $35.46\pm1.13$   &   $ 44.53\pm1.33$   &    $55.81\pm1.49$  &   $64.86 \pm1.35$  \\
          & DeepLink&  $34.61\pm1.69$   & $54.25\pm1.59  $   &   $ 59.12\pm1.29$  &   $ 62.68\pm1.61$  &   $66.78 \pm1.12$   \\
 & \textbf{\textsc{Perfect}} & $\mathbf{44.53}\pm\mathbf{1.23}$ & $\mathbf{61.28}\pm\mathbf{0.93}$ & $\mathbf{67.85}\pm\mathbf{1.11}$ & $\mathbf{70.76}\pm\mathbf{0.68}$ & $\mathbf{72.33}\pm\mathbf{0.73}$\\
        \hline
        \multirow{6}{*}{\tabincell{c}{Twitter\\$\&$\\ Quora}}  
                       & PALE &   $25.36 \pm1.74$  &   $29.43 \pm1.01$  &   $44.72 \pm0.75$   &   $51.03 \pm1.05$ &    $56.80 \pm1.14$  \\  
                       & IONE &   $27.71 \pm0.87$  &   $33.57 \pm1.14$  &   $45.39 \pm1.23$  &   $50.92 \pm0.93$  &    $56.37 \pm1.06$   \\
                      & SNNA &   $31.05 \pm1.36$  &   $34.20 \pm1.45$  &   $39.55 \pm1.75$  &   $52.86 \pm1.66$  &    $59.49 \pm1.47$   \\
                  & MOANA &   $25.85 \pm1.04$  &   $34.59 \pm0.94$  &   $43.12 \pm1.23$  &   $54.07 \pm1.13$ &     $62.48 \pm1.05$   \\
                  & DeepLink&   $36.21 \pm1.33$  &   $53.67 \pm1.29$  &   $57.93 \pm1.01$  &   $61.52 \pm1.21$ &     $66.23 \pm0.86$    \\
      & \textbf{\textsc{Perfect}} &$\mathbf{42.12}\pm\mathbf{0.93}$ & $\mathbf{58.35}\pm\mathbf{0.70}$ & $\mathbf{66.67}\pm\mathbf{0.91}$ & $\mathbf{69.24}\pm\mathbf{0.53}$ & $\mathbf{71.02}\pm\mathbf{0.55}$\\
        \hline
    \end{tabular}
    }
    \vspace{-0.07in}
    \label{userPK}
  \end{table*}
  \begin{table*}[htbp]
  \caption{The MAP of user alignment on DBLP-AMiner and Twitter-Quora datasets (\%)}
  \vspace{-0.07 in}
    \scriptsize
    \centering
  \resizebox{1\textwidth}{!}{
      \begin{tabular}{p{1.2cm}<{\centering}| p{1.5cm}<{\centering} | p{1.7cm}<{\centering}   p{1.7cm}<{\centering} p{1.7cm}<{\centering}  p{1.7cm}<{\centering} p{1.7cm}<{\centering} }
        \hline
        {\textbf{Dataset}} &{\textbf{Method}} & {$\textbf{k=10}$}&  {$\textbf{k=15}$}& {$\textbf{k=20}$}& {$\textbf{k=25}$}& {$\textbf{k=30}$}\\
           \hline
            \multirow{6}{*}{\tabincell{c}{DBLP\\$\&$\\AMiner} }
             & PALE&     $11.13 \pm0.65  $  & $ 11.75\pm0.38$  &  $ 12.67\pm0.29$ &  $ 12.96\pm0.46 $  &   $ 13.16\pm0.50$  \\  
              &IONE &    $12.38 \pm0.37  $  & $ 12.93\pm0.56$  &   $13.76\pm0.62$  & $13.99\pm0.40  $  &   $ 14.22\pm0.48$  \\
              & SNNA&   $13.24\pm0.51  $  & $ 13.66\pm0.69$  &   $ 14.00\pm0.72$  &   $14.63\pm0.78$  &$14.84 \pm0.59$  \\
          & MOANA&   $11.67\pm0.55  $  &  $ 12.43\pm0.43$ &   $12.99\pm0.50$  &  $13.52\pm0.44$  &  $ 13.86 \pm0.48$  \\
          & DeepLink&  $16.45\pm0.60  $ &   $ 18.13\pm0.44$ &   $18.41\pm0.40$   & $ 18.58\pm0.47$  &  $18.73 \pm0.43$  \\
 & \textbf{\textsc{Perfect}} & $\mathbf{18.86}\pm\mathbf{0.47}$ & $\mathbf{20.38}\pm\mathbf{0.37}$ & $\mathbf{20.79}\pm\mathbf{0.36}$ & $\mathbf{20.93}\pm\mathbf{0.21}$ & $\mathbf{20.99}\pm\mathbf{0.22}$\\
        \hline
        \multirow{6}{*}{\tabincell{c}{Twitter\\$\&$\\ Quora}}  
                       & PALE  &    $10.87\pm0.61$  &   $11.24 \pm0.35$  &   $12.19 \pm0.27$  &   $12.49 \pm0.37$ &   $12.71\pm0.39$  \\  
                       & IONE   &   $12.05\pm0.33$  &   $12.56 \pm0.40$  &   $13.28 \pm0.51$   &   $13.53 \pm0.31$  &   $13.74\pm0.42$  \\
                      & SNNA   &   $12.93\pm0.52$  &   $13.23 \pm0.59$  &   $13.57 \pm0.63$   &   $14.22 \pm0.62$ &   $14.48\pm0.57$  \\
                  & MOANA  &   $13.68\pm0.39$  &   $14.47 \pm0.33$   &   $15.01 \pm0.37$   &   $15.52 \pm0.47$   &   $15.84\pm0.37$  \\
                  & DeepLink  &   $16.26\pm0.55$  &   $17.83 \pm0.48$  &   $18.09 \pm0.35$   &   $18.26 \pm0.50$   &   $18.43\pm0.33$  \\
      & \textbf{\textsc{Perfect}} & $\mathbf{17.55}\pm\mathbf{0.38}$ & $\mathbf{18.98}\pm\mathbf{0.25}$ & $\mathbf{19.51}\pm\mathbf{0.31}$ & $\mathbf{19.64}\pm\mathbf{0.18}$ & $\mathbf{19.70}\pm\mathbf{0.20}$\\
        \hline
    \end{tabular}
    }
    \vspace{-0.1in}
    \label{userMAPK}
  \end{table*}

\subsubsection{Experimental Results and Discussions}
First, we evaluate the performance  under different overlap rates $\eta$.
The overlap rate $\eta$ is defined as $2|\mathcal A|/(N^s+N^t)$, where $|\mathcal A|$, $N^s$ and $N^t$ are the number of anchor users, source network users and target network users, respectively. 
An  $\eta-$overlap dataset is generated by randomly deleting users according to the overlap rate from the dataset.
Fixing embedding dimension $d=64$, we report results under $\eta=\{20,  30, 40, 50, 60\}(\%)$ in terms of $Accuracy$ and $Quality$ on both datasets in Fig. \ref{comAlignEta}.
Second, we discuss the effect of embedding dimension $d$. 
Specifically, embedding dimension $d$ takes different values in $[8, 16, 32, 64, 128]$ and we report results under $\eta=60\%$ in terms of $Accuracy$ and $Quality$ in Fig. \ref{comAlignDi}. 
Note that, the alignment threshold $\tau$ of $Accuracy$ is set to $60\%$ in the experiments above. 
Next, we study the parameter sensitivity of $\tau$ and report results under $d=64$ and  $\eta=60\%$  in Fig. \ref{comAlignTau}. 

Finally, we summarize our findings and discuss the reasons: 
\begin{itemize}
\item \emph{\textsc{Perfect} consistently outperforms its competitors.}
The reason is that \textsc{Perfect} enjoys the strengths of hyperbolic space and the unified optimization closing the loop of community alignment and user alignment.
\item \emph{The models of hyperbolic space consistently beat the Euclidean ones as shown in both Figs. \ref{comAlignEta} and \ref{comAlignDi}.} 
Moreover, we obtain better performance on the dataset of higher hyperbolicity. (Refer to Section \ref{sec:why} and Table \ref{stat}.) 
The reason is that hyperbolic space better matches the inherent hyperbolicity of these datasets than the Euclid, 
and hyperbolic space benefits community alignment.
\item \emph{The models of unified optimization perform better in general as shown in both Figs. \ref{comAlignEta} and \ref{comAlignDi}.}
The reason is that, in a unified optimization, community alignment and user alignment benefit each other, while community alignment neglects the effect of user alignment and vice versa in the two-step methods.
\item \emph{The proposed hyperbolic model, \textsc{Perfect}, with low-dimensional embeddings (e.g., 16) outperforms Euclidean ones with high-dimensional embeddings (e.g., 128) as shown in Fig. 4. }
The reason is that, well suited for networks with latent hierarchy, \textsc{Perfect} generates faithful embeddings with a few dimensions in hyperbolic space while it is not true for Euclidean ones.
\item \emph{\textsc{Perfect} shows better robustness regarding threshold $\tau$ as shown in Fig. 5.}
The reason is that, in \textsc{Perfect}, there tends to be more anchor users among aligned communities, \ie, better results regardless of $\tau$.
\end{itemize}

\subsection{Experiments on User Alignment}\label{sec:exp-user}
In this part, we will illustrate the learning performance of \textsc{Perfect} compared against the comparison methods on user alignment.
\subsubsection{Comparison Methods}
We chose several state-of-the-art methods on user alignment as follows:
\begin{itemize}
\item \emph{IONE} \cite{liu2016aligning}: It embeds social networks together with anchor links in the Euclidean subspace to align users.
\item \emph{PALE} \cite{man2016predict}: It first embeds each network and then matches users via the Euclidean metric.
\item \emph{DeepLink} \cite{zhou2018deeplink}: It leverages dual learning to refine the Euclidean subspace where network alignment is performed. 
\item \emph{SNNA} \cite{li2019adversarial}: It proposes a weakly-supervised adversarial learning method for alignment from the distribution level. 
\item \emph{MOANA} \cite{zhang19multi}: It introduces a coarsen-align-interpolate method via matrix analysis to find node correspondence.
\end{itemize}

\begin{figure}
\centering 
\subfigure[DBLP-AMiner: Precision$@30$]{
\includegraphics[width=0.49\linewidth]{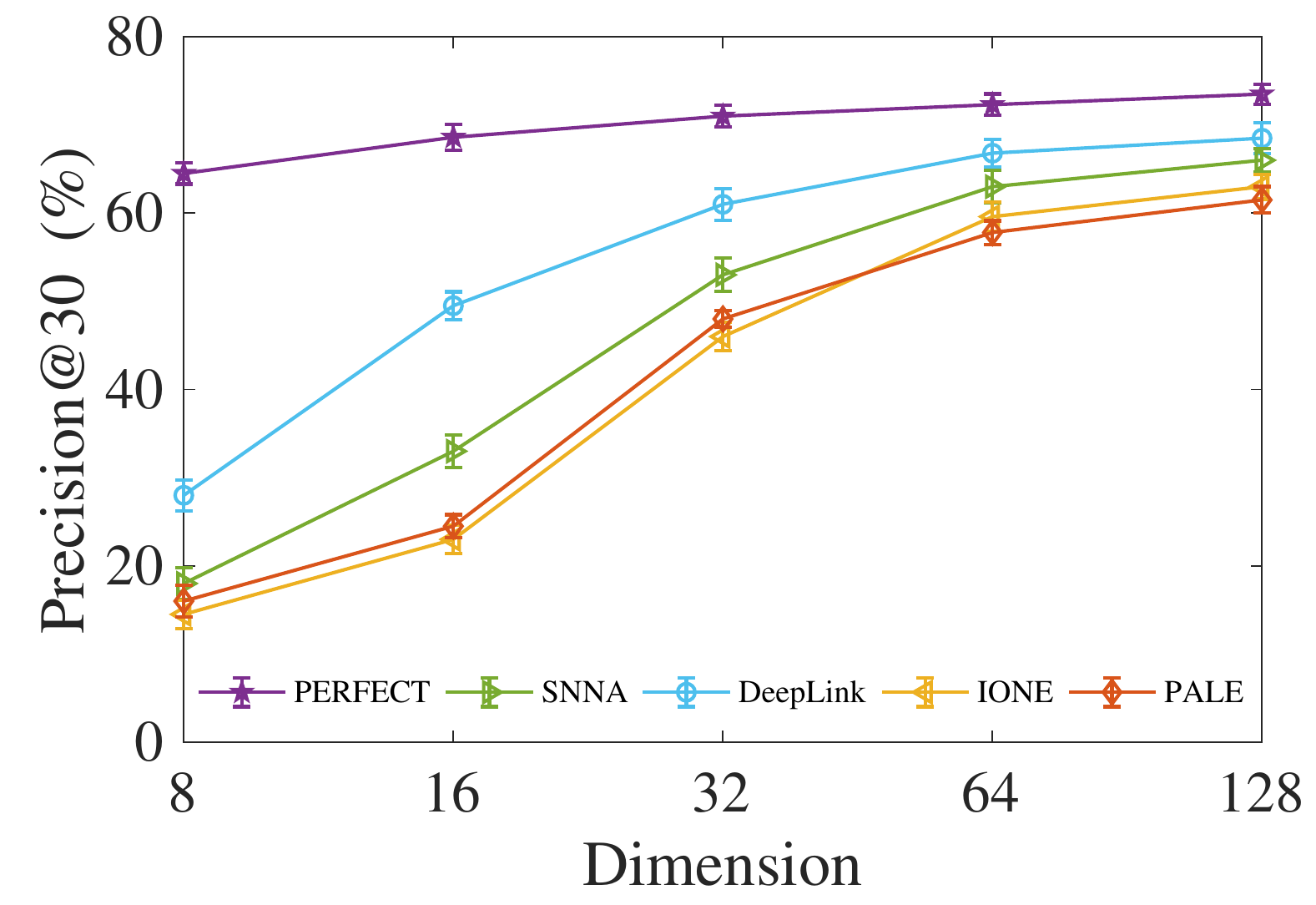}}
\hspace{-0.036\linewidth}
\subfigure[Twitter-Quora: Precision$@30$]{
\includegraphics[width=0.49\linewidth]{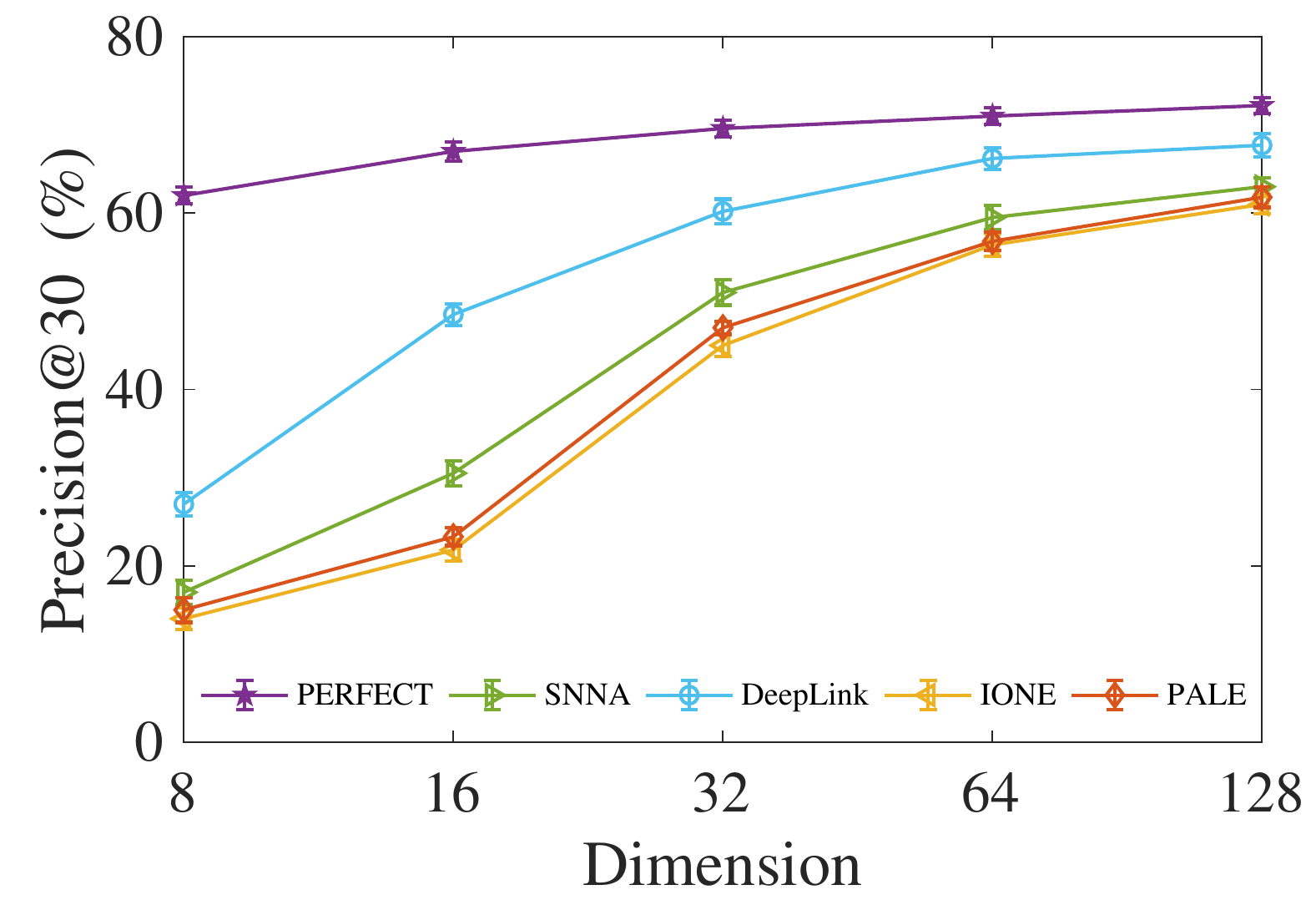}}
\vfill
\subfigure[DBLP-AMiner: MAP$@30$]{
\includegraphics[width=0.49\linewidth]{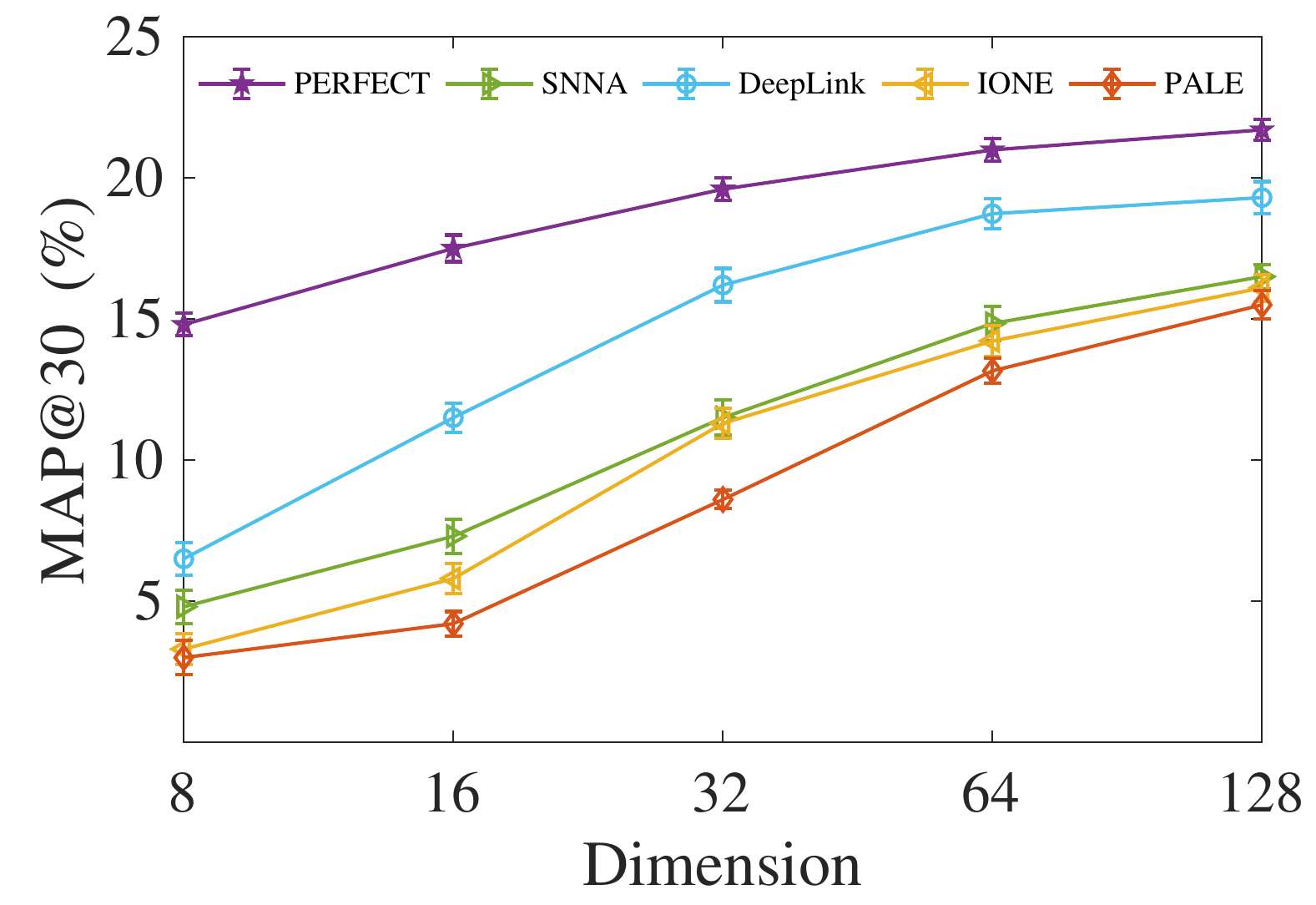}}
\hspace{-0.036\linewidth}
\subfigure[Twitter-Quora: MAP$@30$]{
\includegraphics[width=0.49\linewidth]{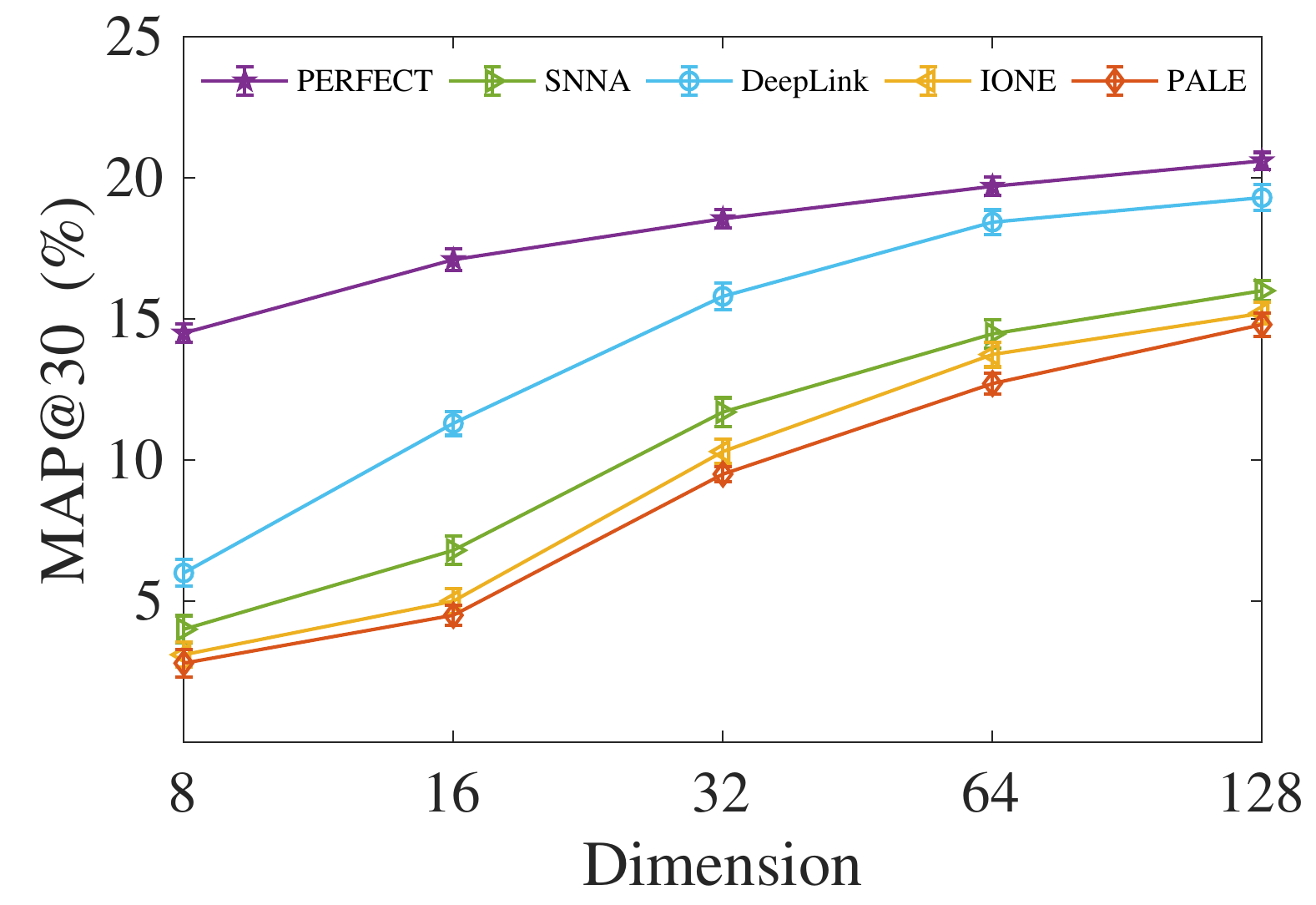}}
\vspace{-0.15in}
\caption{User alignment under different dimensions}
\vspace{-0.25in}
\label{userAlignDi}
\end{figure}

\begin{figure*} 
\centering 
\subfigure[The embeddings of DBLP]{
\includegraphics[width=0.25\linewidth]{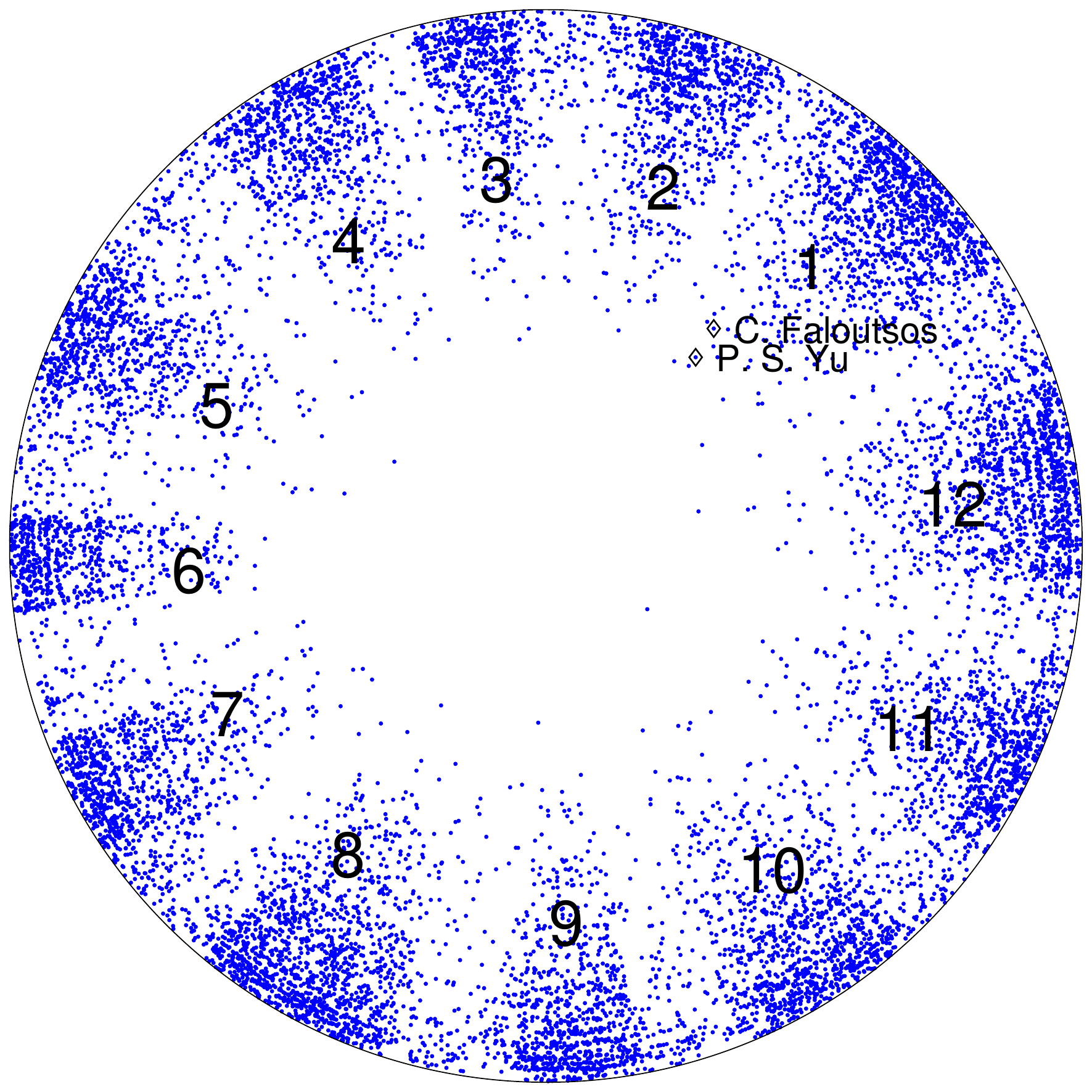}}
\hspace{0.025\linewidth}
\subfigure[The common Poincar\'{e} disk]{
\includegraphics[width=0.25\linewidth]{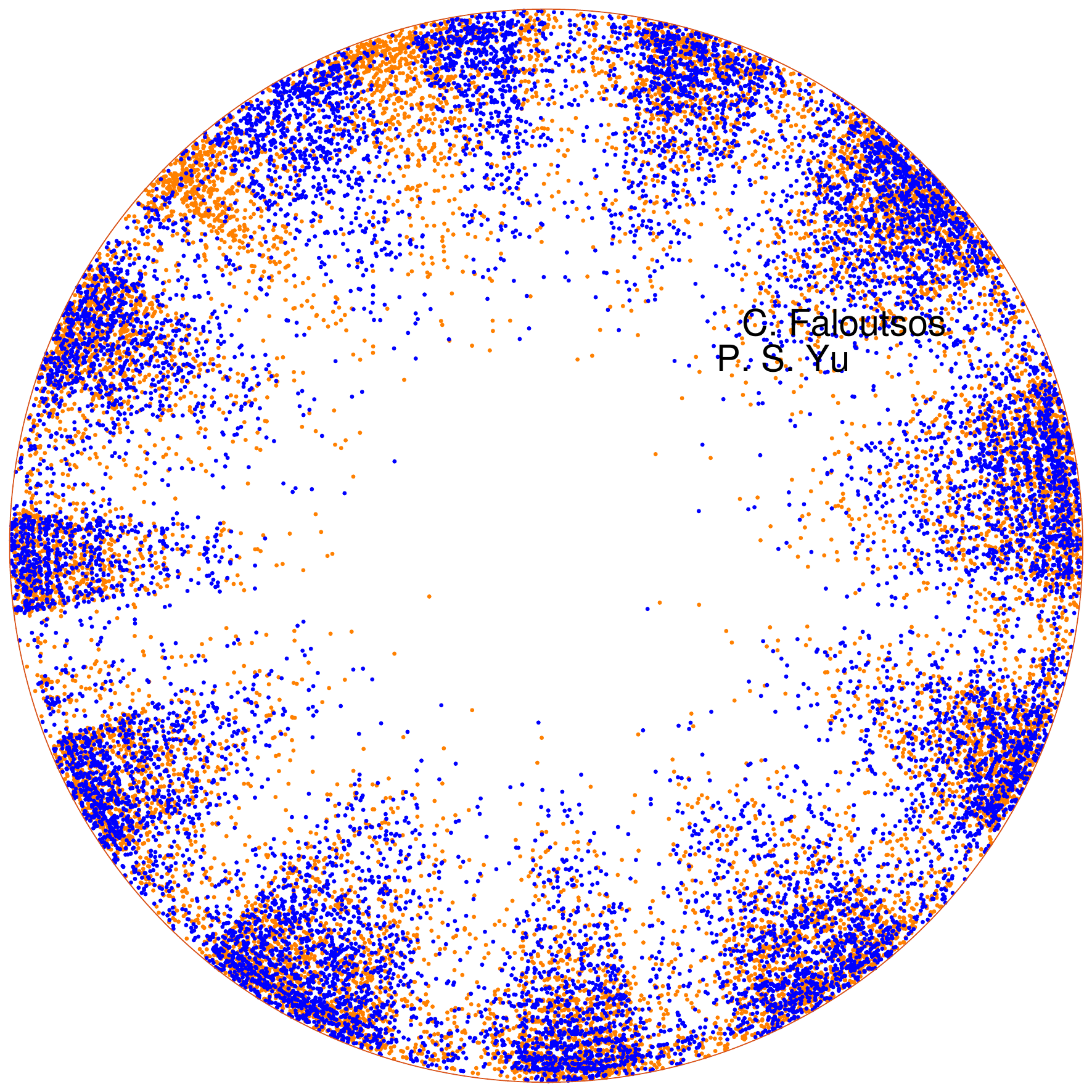}}
\hspace{0.025\linewidth}
\subfigure[The embeddings of AMiner]{
\includegraphics[width=0.25\linewidth]{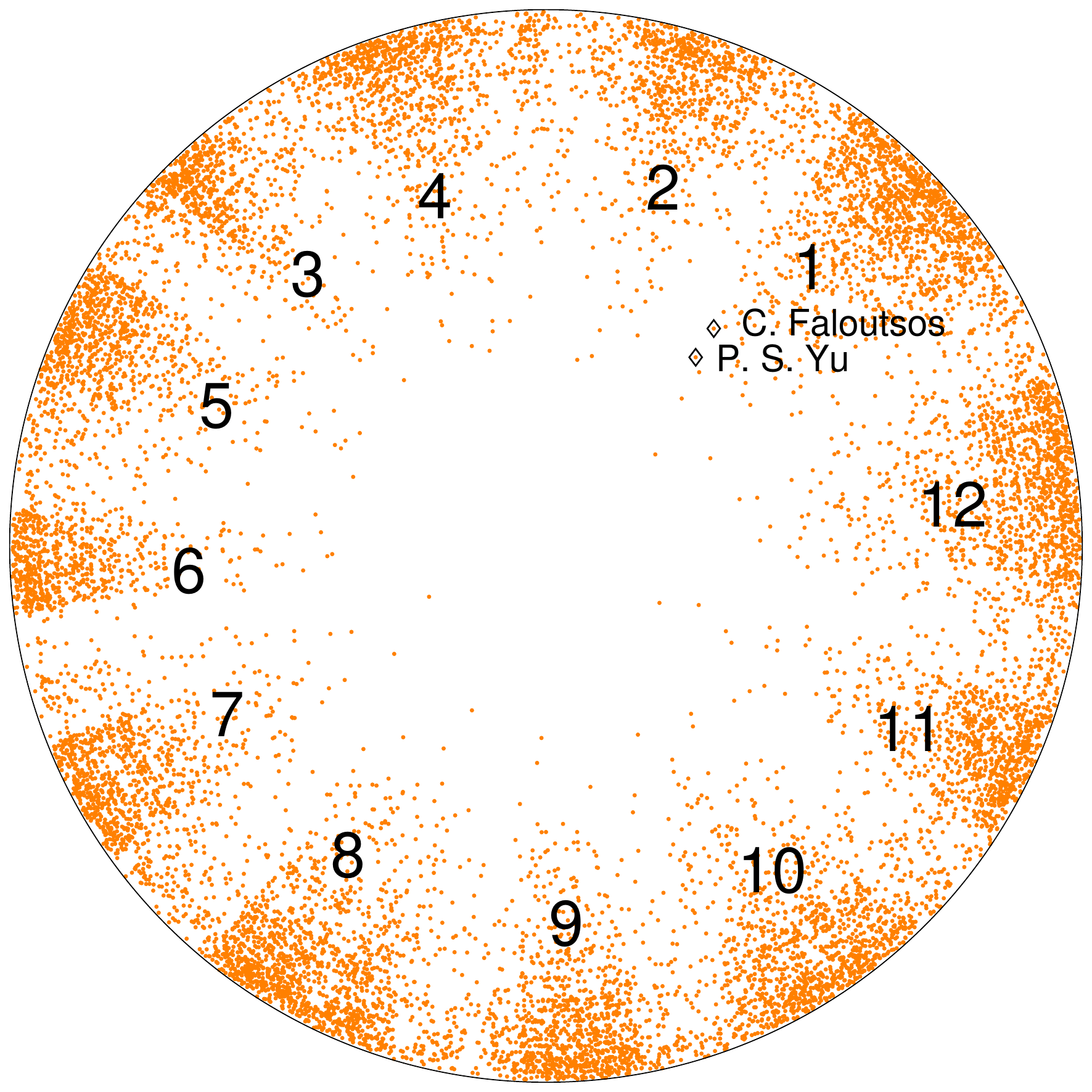}}
\hspace{0.025\linewidth}
\subfigure{
\includegraphics[width=0.1\linewidth]{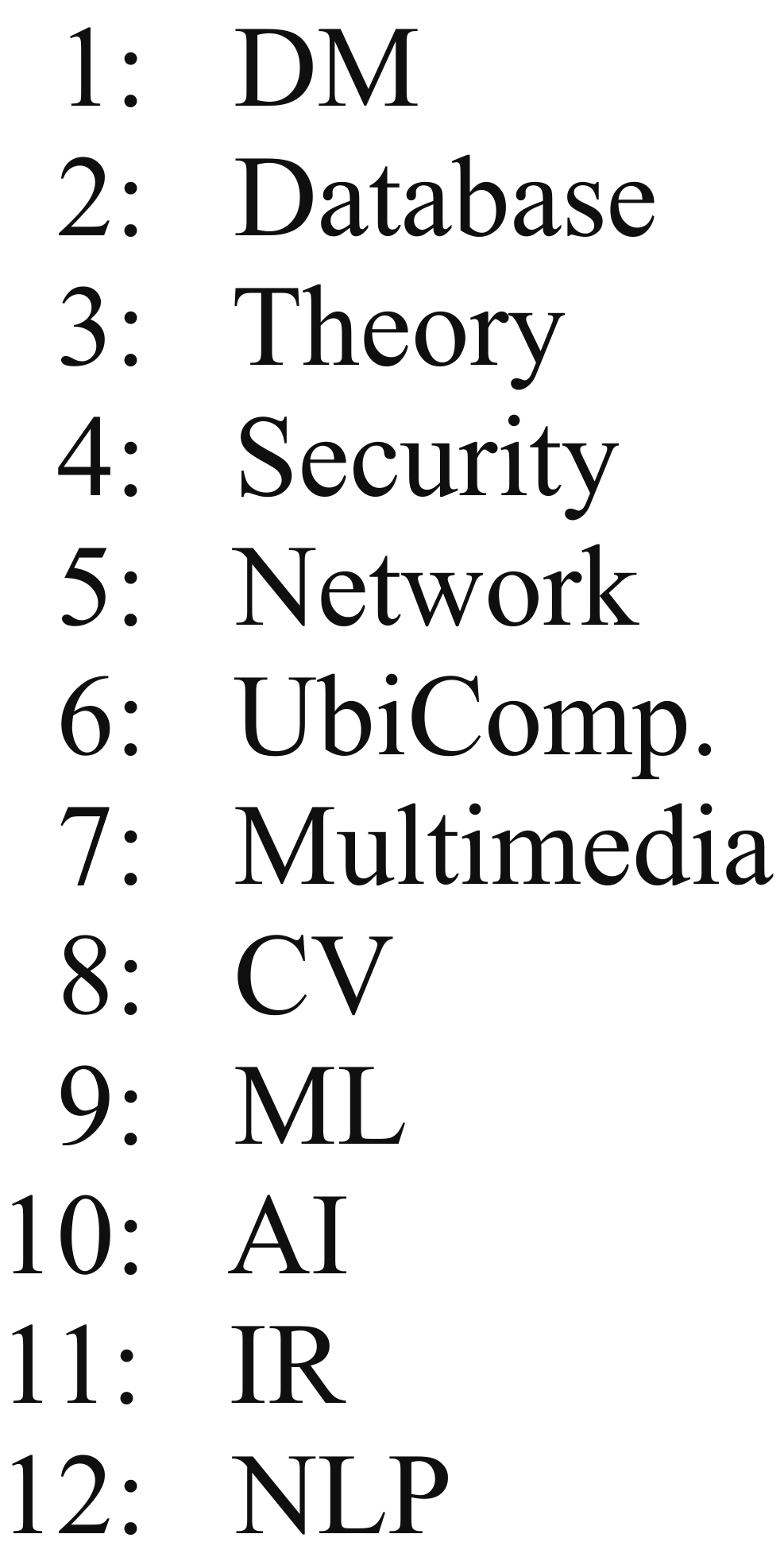}}
\vspace{-0.1in}
\caption{The common Poincar\'{e} disk of DBLP-AMiner dataset}
\vspace{-0.2in}
\label{case}
\end{figure*}

\subsubsection{Evaluation Metric}
We employ the widely used $Precision@K$ and $MAP@K$ as the evaluation metric.
\subsubsection{Experimental Results and Discussions}
First, we evaluate the performance of user alignment in terms of $Precision@k$ and $MAP@k$ on both datasets, 
whose experimental results are shown in Table \ref{userPK} and Table \ref{userMAPK}, respectively. 
Specifically, $k$ takes different values in $[10, 15, ..., 30]$ and embedding dimension is set to $64$.
Our findings are two-fold:

\noindent \textbf{(1)} \emph{\textsc{Perfect} consistently outperforms its comparison methods. }
The reasons lie in that, 
besides the structural information encoded in typical embedding methods, \eg, DeepLink, IONE, PALE, 
user embeddings in \textsc{Perfect} incorporate the community structure explicitly and latent hierarchy among users in hyperbolic space implicitly, 
and thus own more discriminative information for user alignment.
Additionally, this demonstrates that  hyperbolic space benefits the user alignment.

\noindent \textbf{(2)} \emph{The performance gain of \textsc{Perfect} is correlated to the network hyperbolicity.}
The proposed approach achieves higher performance on the dataset of higher hyperbolicity (DBLP-AMiner), 
and beats Euclidean methods on both datasets.

Second, we discuss the effect of embedding dimension $d$.
We take different values in $[8, 16, 32, 64, 128]$, and report the performance of user alignment in terms of $Precision@30$ and $MAP@30$ on both datasets in Fig. \ref{userAlignDi}.
We omit the performance of  MOANA as it is not an embedding method.
We find that \emph{as shown in Fig. \ref{userAlignDi}, \textsc{Perfect} consistently outperforms its competitors in all dimensions, and obtains dramatic performance gain with low-dimensional embeddings.}

\vspace{-0.05in}
\subsection{Case Study} \label{sec:exp-case}
In this part, we will give a case study on DBLP-AMiner dataset to demonstrate the performance of \textsc{Perfect} on both user alignment and community alignment.

We visualize the common Poincar\'{e} disk of \emph{\textsc{Perfect}} in Fig. \ref{case} (b) by setting $d=2$.
We filter the embeddings of each network for clarity.
The embeddings of  DBLP and AMiner are plotted in Fig. \ref{case} (a) and (c), respectively, where communities are labeled as illustrated in the legend.
In this common Poincar\'{e} disk, we have three main findings:

\noindent \textbf{(1)} \emph{The latent hierarchy among users is generally preserved in the two-dimensional embeddings.}
We find that authors of high impact, \eg, Philip S. Yu and Christos Faloutsos, are positioned closer to the origin, while those of relative low impact are pushed to the boundary of the disk.

\noindent \textbf{(2)} \emph{User embeddings of the same user in different networks locate closely.}
Zoom in the community of data mining. The user embedding of  Philip S. Yu in DBLP, shown in Fig. \ref{case} (a), is nearly the same as that in AMiner, Fig. \ref{case} (c). Thus, user alignment is easy to be inferred via \textsc{Perfect}.

\noindent \textbf{(3)} \emph{Centripetal regions of corresponding communities are naturally aligned.}
We find that user embeddings of the same community cluster into a centripetal region, which is also reported in the study \cite{muscoloni2017machine}, and it is obvious that most of centripetal regions of the same research area are aligned.
That is, both users and communities are aligned in the common Poincar\'{e} disk, verifying the basic idea of our approach.

\section{Related Works}
We briefly summarize the related works in following areas:

\emph{\textbf{Network embedding}} is to map the nodes of a network into a vector space \cite{cui2018survey,zhang2018network}.
Here, we roughly classify the literature by the embedding space. 
Most existing studies \cite{perozzi2014deepwalk,tang2015line,grover2016node2vec} explicitly or implicitly work with the Euclidean space. 
Some studies \cite{tu2018deep,pan2018age} focus on exploiting in the static or temporal network structure, while others \cite{huang2017label,peng18www} attempt to incorporate other  attributes.
However, is  Euclidean space the appropriate embedding space?
Recent advances uncover the hyperbolicity of some real-world networks, and researchers attempt to facilitate network embedding in hyperbolic space.
Some studies \cite{nickel2017poincare,wang2019hyperbolic} suggest the superiority of hyperbolic geometry for network embedding.
Recently, some neural networks on graphs \cite{chami2019hyperbolic,liu2019hyperbolic,gulcehre2018hyperbolic} learn node embeddings underpinned by hyperbolic space.
Distinguishing from these methods, \textsc{Perfect} is tailored for network alignment.





\emph{\textbf{User alignment}}, or anchor link prediction, is to align users across different social networks according to the underlying identity.
To address this problem, some studies \cite{liu2014hydra,mu2016user} leverage attribute information, such as screen name pattern and user behaviors, to discover the identity consistency, 
while some studies \cite{man2016predict} exploit network topology to link user identities. 
Moreover, there exist studies \cite{Zhong2018CoLink} considering both topology and attribute information.
The study \cite{shu2017user} gives a comprehensive survey.
Recently, 
HGANE \cite{jiao2019collective} incorporates attentive mechanism to induce the common subspace.
GAlign \cite{hzyin2020galign} performs network alignment in an unsupervised way.
Note that, similar to our prior works \cite{sun2019dna,sun2018ijcai,sun2020tkde,ren2020ijcai}, all of methods in the literature work with the Euclidean space, while \textsc{Perfect} works with hyperbolic space.
Additionally, beyond user alignment, \textsc{Perfect} jointly considers community alignment in a unified approach. 


\emph{\textbf{Communities}} play a fundamental role in social network analysis.
To the best of our knowledge, most existing studies \cite{abbe2017community} focus on community discovery in an isolated social network.
Various types of methods have been proposed, such as modularity optimization \cite{newman2006modularity} and spectral algorithms \cite{chin2015stochastic}.
Generative methods are often explored as well, such as community affiliation models \cite{yang2013overlapping}, model-based clustering and GAN \cite{jia2019communitygan}.
Actually, social networks are correlated and partially aligned on their common users. 
Some studies \cite{zhang2015community} facilitate community discovery with the information of its counterpart network.
Recently, the study \cite{chen2017community} considers the community structure to facilitate user alignment across social networks.
Different from these studies, we for the first time close the loop of community alignment and user alignment so that they mutually enhanced each other.

\section{Conclusion}
In this paper, 
we present a novel hyperbolic optimization framework, namely \textsc{Perfect},
to jointly align users and communities in hyperbolic space.
To address the optimization of \textsc{Perfect}, we propose a novel alternating Riemannian optimization algorithm with solid theoretical analyses 
so that user alignment and community alignment benefit from each other.
Extensive experiments show the superiority of \textsc{Perfect} in both user alignment and community alignment.

\section*{Acknowledgment} 
This work was supported in part by: National Natural Science Foundation under Grants U1936103 and 61921003, National Key Research and Development Program of China under Grant 2018YFB1003804, Fundamental Research Funds for the Central Universities 2019XD11,
Natural Science Foundation under Grants III-1526499, III-1763325, III-1909323, SaTC-1930941, IIS-1763365, 
and by Florida State University.

\scriptsize
\bibliographystyle{IEEEtran}
\bibliography{main}

\end{document}